\documentclass{article}

\usepackage[english]{babel}

\usepackage[letterpaper,top=2cm,bottom=2cm,left=3cm,right=3cm,marginparwidth=1.75cm]{geometry}

\usepackage{amsmath}
\usepackage{graphicx}
\usepackage{multicol}
\usepackage{enumitem}
\usepackage{cite}
\usepackage{amssymb}

\usepackage[colorlinks=true, allcolors=blue]{hyperref}
\usepackage{authblk}

\title{\bf New avenues for the neutrino dipole portal exploration at the energy frontier}
\author{Natascia Vignaroli}
\affil{Dipartimento di Matematica e Fisica, Università del Salento, and Istituto Nazionale di Fisica
Nucleare, Sezione di Lecce, I-73100 Lecce, Italy}

\begin{document}
\maketitle

\begin{abstract}

We present a comprehensive and gauge-invariant study of the neutrino dipole portal at the energy frontier. Assuming negligible active–sterile mixing, we analyze sterile neutrino production via dimension-6 dipole operators coupling to the electroweak field strengths. The analysis incorporates both single- and double-gauge-boson effective interactions. We investigate novel collider signatures at the HL-LHC, the FCC-hh—studied here in this context for the first time—and a 10 TeV muon collider. Particular emphasis is placed on electroweak boson-initiated processes, which dominate in the high-mass regime above $\sim$1 TeV. At the muon collider, these VBF-like topologies 
enable production even when the dipole couples to non-muonic flavors, offering a unique and sensitive probe for different flavor scenarios.  
We derive sensitivity projections for various theoretical 
benchmarks, reaching dipole couplings down to $d_\gamma\sim 6\times10^{-7}$ GeV$^{-1}$ at FCC-hh and $d_\gamma\sim 2\times10^{-7}$ GeV$^{-1}$ at the muon collider.

\end{abstract}
\newpage

\tableofcontents

\section{Introduction}

The observation of neutrino oscillations provides unambiguous evidence that neutrinos are massive, implying the existence of physics beyond the Standard Model (SM). A minimal extension involves the addition of gauge-singlet right-handed neutrinos, often referred to as \emph{sterile neutrinos} ($N$), which can couple to the SM via Yukawa interactions with lepton doublets and the Higgs field. These interactions, together with Majorana mass terms for the sterile states, can generate active neutrino masses through the seesaw mechanism \cite{Minkowski:1977sc, Gell-Mann:1979vob, Yanagida:1979as, Mohapatra:1979ia}. Furthermore, sterile neutrinos may play a key role in explaining the baryon asymmetry of the Universe through leptogenesis~\cite{Fukugita:1986hr,Akhmedov:1998qx}.

In this framework, the sterile neutrinos acquire electroweak interactions suppressed by the active--sterile mixing angle $ \theta \sim \sqrt{m_\nu/m_N} $, which becomes small for sterile neutrino masses above the electroweak scale. As a consequence, obtaining evidence for the existence of sterile neutrinos based on mixing-induced interactions is challenging in this mass regime. However, if additional interactions are present, sterile neutrinos may remain experimentally accessible even when the mixing is negligible.

Among the possible new interactions, \emph{neutrino dipole portal} offers an attractive minimal and phenomenologically rich alternative. This portal connects active and sterile neutrinos via effective operators involving the field-strength tensors of the SM electroweak gauge symmetry. 
These interactions can dominate over the mixing-induced ones in scenarios where the active--sterile mixing is strongly suppressed or vanishing~\cite{Bertuzzo:2024eds}. In the electroweak broken phase, such dipole operators generate magnetic-type couplings to the photon, $ Z $, and $ W $ bosons. 

Previous studies have analyzed the phenomenology of the dipole portal in both low-energy and high-energy regimes. For sterile neutrinos with sub-GeV masses, strong bounds arise from astrophysical observations, beam-dump experiments, and rare-meson decays~\cite{Bertuzzo:2024eds, Brdar:2020quo, Magill:2018jla}. 
At higher masses, collider-based searches become relevant. Several analyses have considered sterile neutrino production and decay via the photon dipole interaction, 
both at the LHC and at lepton colliders such as LEP~\cite{Delgado:2022fea}
or future $ e^+ e^- $ facilities such as CEPC~\cite{Zhang:2023nxy}. 
Projections for LHC in its high luminosity phase (HL-LHC) have also been derived using searches for long-lived particles with displaced tracks~\cite{Jodlowski:2020vhr,
Delgado:2022fea, Duarte:2023tdw, Ovchynnikov:2023wgg, Beltran:2024twr} and prompt $ N \to \gamma \nu$ decays~\cite{Magill:2018jla}.

It is essential to consider that a complete and consistent formulation of the dipole portal at collider energies requires the use of dimension-6 operators invariant under the full SU(2)$_w$ × U(1)$_Y$ SM gauge symmetry. After electroweak symmetry breaking, these operators generate correlated dipole couplings to the $ \gamma $, $ Z $, and $ W $ bosons. This structure implies the existence of both neutral- and charged-current processes, and enhances the variety of possible collider signatures. Notably, charged-current channels such as $ N \rightarrow W \ell $ can lead to clean leptonic final states, and radiation of electroweak bosons from the initial state becomes increasingly important at high energies. These features have recently been explored at a future muon collider~\cite{Barducci:2024kig, Brdar:2025iua}.

In this work, we present a \emph{comprehensive and fully gauge-invariant analysis} of the neutrino-dipole portal at collider experiments, including not only the usual single-gauge-boson effective interactions, but also \emph{genuine two-boson terms} that arise from the SU(2)\(_w\)-invariant operator and are necessary for a correct treatment of neutrino-dipole portal processes at high energy.
We assume a sterile neutrino with negligible active--sterile mixing and dominant dipole-mediated interactions. 

The analysis covers different collider scenarios. We will start by exploring the reach of the upcoming High-Luminosity LHC (HL-LHC) program, for which we will investigate  
the range $m_N \gtrsim$ 100 GeV, focusing in particular on the decays $N \to W l$.
We will then consider experiments at the energy frontier, in particular those supported by the recent European and US strategies for particle physics: a future circular collider (FCC)~\cite{FCC:2025lpp, FCC:2018byv} and a futuristic muon collider~\cite{Accettura:2023ked}. 
\emph{For the first time, we will investigate the sensitivity of the FCC-hh}~\cite{FCC:2018vvp} and reconsider the sensitivity of a muon collider, recently explored in \cite{Barducci:2024kig, Brdar:2025iua}. The latter have shown that a muon collider would reach expected sensitivities of the order $\mathcal{O}(10^{-7})$ GeV$^{-1}$, which would be the best ever achieved in the region $m_{N}\gtrsim1$ TeV.
The sensitivity of future colliders to SM neutrino-dipole moments has instead been recently analyzed in~\cite{Frigerio:2024jlh}.

A key focus of the present work is the role of electroweak boson radiation from the initial state, which gives rise to \emph{sterile neutrino production via} scattering or fusion of weak bosons, which we concisely dub \emph{vector-boson-fusion (VBF)}. Although VBF channels have not been systematically analyzed in the context of the dipole portal (a recent study~\cite{Dehghani:2025xkd} has shown their potential for the search for heavy Majorana neutrinos at muon colliders in the presence of significant mixing with the SM), we demonstrate that they provide very high sensitivity at high energies and large sterile neutrino masses. In particular, we present the \emph{first dedicated analysis of these production modes at a muon collider}, showing that they allow sterile neutrinos to be efficiently produced \emph{even when the dipole operator couples to flavors other than the muon}.

The remainder of the paper is structured as follows. In Section~\ref{sec:theory}, we introduce the theoretical framework, including the relevant gauge-invariant effective operators and their low-energy projections. 
After discussing sterile neutrino decay modes (\ref{subsec:decays}),
in Section~\ref{sec:hadron} we present the collider analysis for hadronic machines: we will first discuss production channels (Sect.~\ref{sec:production-hadron}) and then we will evaluate the sensitivity reach at the HL-LHC (Sect.~\ref{sec:hl-lhc}) and at the FCC-hh (Sect.~\ref{sec:fcc}). Section~\ref{sec:mucol} is devoted to muon collider analysis: we describe the dominant production channels in Sect.~\ref{sec:production-mucol}, and present search strategies and sensitivity projections in VBF channels 
in Sect.~\ref{sec:mucol-sensitivities}. 
Our conclusions are summarized in Section~\ref{sec:conclusions}.


\section{Theoretical framework}\label{sec:theory}

In this study, we work under the hypothesis that neutrino-transition dipole moments constitute the dominant active-sterile portal interactions. We therefore assume negligible mixing between sterile and active neutrinos.

Dipole portal moments arise from the following 6-dimensional operators, which are generated by some new physics 
above the EW scale and which must respect the full $SU(2)_w\times U(1)_Y$ gauge symmetry of the SM:
\begin{align}
\mathcal{L} & \supset \;  \mathcal{C}^i_B \mathcal{O}^i_B + \mathcal{C}^i_W \mathcal{O}^i_W + h.c. \\ \nonumber
& =   \mathcal{C}^i_B g^\prime\bar{L}^i B_{\mu\nu}\tilde{H} \sigma^{\mu\nu} N + \mathcal{C}^i_W g \bar{L}^i W^a_{\mu\nu} \tau^a \tilde{H} \sigma^{\mu\nu} N + h.c. \; ,
\end{align}
where $i$ is a flavor index,  $g^\prime=e \cos\theta_w$ and $g=e \sin\theta_w$ are the $U(1)_Y$ and $SU(2)_w$ gauge couplings, $\tau^a=\sigma^a/2$, with $\sigma^a$ the Pauli matrices, $2\sigma^{\mu\nu}=i[\gamma^\mu,\gamma^\nu]$, $B_{\mu\nu}^a$ and $W_{\mu\nu}$ are the gauge field strengths of $U(1)_Y$ and $SU(2)_w$, $\tilde{H}=i \sigma^2 H^\star$ is the conjugate Higgs field, $L$ is the lepton doublet and $N$ the sterile neutrino, assumed to be of Majorana nature.

After spontaneous symmetry breaking of the Higgs, one obtains the following 5-dimensional effective interactions: 
\begin{align}\label{eq:dipoles}
\begin{split}
\mathcal{L} \supset & \; d^i_{\gamma} \bar{\nu}^i_L \sigma^{\mu\nu} N F_{\mu\nu}+  d^i_{Z} \bar{\nu}^i_L \sigma^{\mu\nu} N Z_{\mu\nu} +  d^i_{W} \bar{\ell}^i_L \sigma^{\mu\nu} N W^{-}_{\mu\nu} \\
& - d^i_W 2 i g \, \bar{\ell}^i_L \sigma^{\mu\nu} N\, W^{-}_\mu (s_{\theta_w} A_\nu + c_{\theta_w} Z_\nu) +  d^i_W \sqrt{2} i g \, \bar{\nu}^i_L \sigma^{\mu\nu} N \, W^{-}_\mu W^{+}_\nu + h.c. \; ,
\end{split}
\end{align}
 where $F_{\mu\nu}$ and $Z_{\mu\nu}$ are, respectively, the photon and the $Z$ boson field strengths, $s_{\theta_w}$ ($c_{\theta_w}$) shortly denotes the sin (cosin) of the Weinberg angle, and $W^{-}_{\mu\nu}\equiv\partial_\mu W^{-}_\nu - \partial_\nu W^{-}_\mu$. Note the presence of interaction terms between fermionic currents and two gauge bosons, which arise from the operator $\mathcal{O}_W$. These terms have been typically neglected in the previous literature \cite{Magill:2018jla, Barducci:2024kig}~\footnote{I thank Vedran Brdar for clarifying that these terms were correctly included in the calculations performed in paper~\cite{Brdar:2025iua}, even though they were not explicitly written in the initial version of their manuscript. }, but play a crucial role in our study at the frontier of energy, since they are responsible for a dominant contribution to $N$ production through VBF.  

The dipole couplings in the broken phase are matched to the Wilson coefficients in the unbroken phase according to
\begin{align}\label{eq:match}
\begin{split}
&d^i_{\gamma}=\frac{e v}{\sqrt{2}} \left(\mathcal{C}^i_B +  \frac{1}{2} \mathcal{C}^i_W\right) \\ 
&d^i_{W}=\frac{e v}{2 s_{\theta_w}} \mathcal{C}^i_W \\
&d^i_{Z}=\frac{ev}{\sqrt{2}} \left( -\tan\theta_w\mathcal{C}^i_B +  \frac{\cot\theta_w}{2} \mathcal{C}^i_W\right)\; ,
\end{split}
\end{align}
where $v=246$ GeV. \footnote{We find a discrepancy of a factor $1/\sqrt{2}$ with the equation reported in version 1 of manuscript \cite{Brdar:2025iua} for $d_W$ (even taking into account the different definition of $\mathcal{C}_W$), while we agree with other previous literature \cite{Magill:2018jla, Barducci:2024kig}. }
It is clear that the dipole couplings are not independent but intercorrelated through the coefficients $\mathcal{C}_B$ and $\mathcal{C}_W$. The correlation leads to an involved phenomenology that includes interference effects between various contributions. To properly examine the processes induced by a neutrino-dipole portal, we will, therefore, select theoretical benchmarks.
In this study, we will specifically analyze the following:
\begin{multicols}{2}
    \begin{enumerate}[label=(\Alph*)]
     \item $\mathcal{C}_B=0$ 
        \item $\mathcal{C}_W=\frac{2 s_{\theta_w} \tan\theta_w}{c_{\theta_w}-\sqrt{2}} \,\mathcal{C}_B \approx -\mathcal{C}_B \, \Leftrightarrow d_W=d_Z$
        \item $\mathcal{C}_W=2 \tan^2\theta_w\, \mathcal{C}_B\approx 0.57 \,\mathcal{C}_B$ $\Leftrightarrow d_Z=0$
        \item $\mathcal{C}_W=0$ $\Leftrightarrow d_W=0$
    \end{enumerate}
    \end{multicols}

As will become clear in the following sections, these scenarios cover a wide range of possibilities for sterile neutrino phenomenology at colliders. For simplicity, we will assume that $\mathcal{C}_B$ and $\mathcal{C}_W$ are real.\footnote{In the more general case of complex coefficients, possible implications for CP violation are discussed for example in~\cite{Balaji:2020oig}.}

\subsection{Sterile neutrino decays}\label{subsec:decays}

The dipole interactions in Eq.~\eqref{eq:dipoles} allow for different production channels for the sterile neutrino at colliders, which will be examined in the next sections case by case depending on the type of collider, and lead to the three dominant $N$ decays, $N\to \gamma \nu$, $N \to W \ell$, $N \to Z \nu$, with the following rates:
\begin{align}\label{eq:decays}
\begin{split}
& \Gamma_{N \to \nu^i \gamma} = d^{i \, 2}_\gamma \frac{m^3_N}{2\pi}\\ 
&\Gamma_{N \to \ell^{i\, \pm} W^\mp} = d^{i \, 2}_W \frac{m^3_N}{2\pi} \left( 1 - \frac{m^2_W}{m^2_N} \right)^2 \left( 1 + \frac{m^2_W}{2 m^2_N}\right) \\
&  \Gamma_{N \to \nu^i Z} = d^{i \, 2}_Z \frac{m^3_N}{2\pi} \left( 1 - \frac{m^2_Z}{m^2_N} \right)^2 \left( 1 + \frac{m^2_Z}{2 m^2_N}\right)\; ,
\end{split}
\end{align}
where we have neglected the lepton masses, since in our study we are interested in a regime $m_N \gg m_\ell$. Dipole interactions with two gauge bosons (second line in Eq.~\eqref{eq:dipoles}) allow in principle also 3-body $N$ decays. However, the corresponding rates are phase-space suppressed and negligible. 

\begin{figure}[tbp]
\centering
 \includegraphics[scale=0.5]{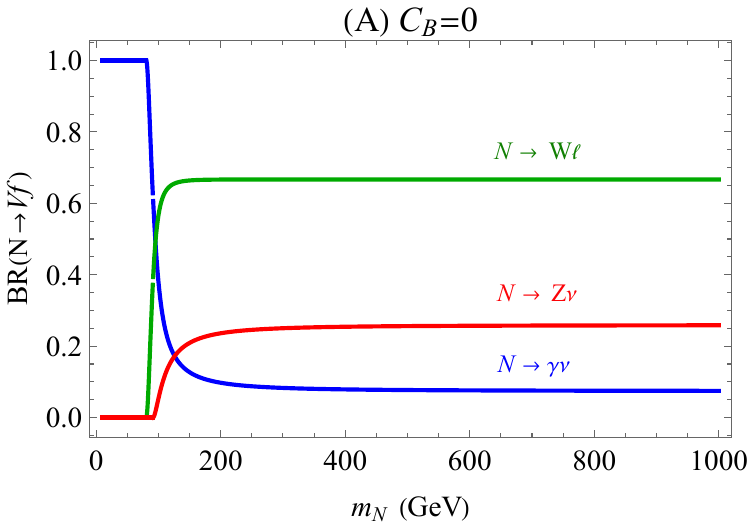} \hspace{0.2cm}\includegraphics[scale=0.5]{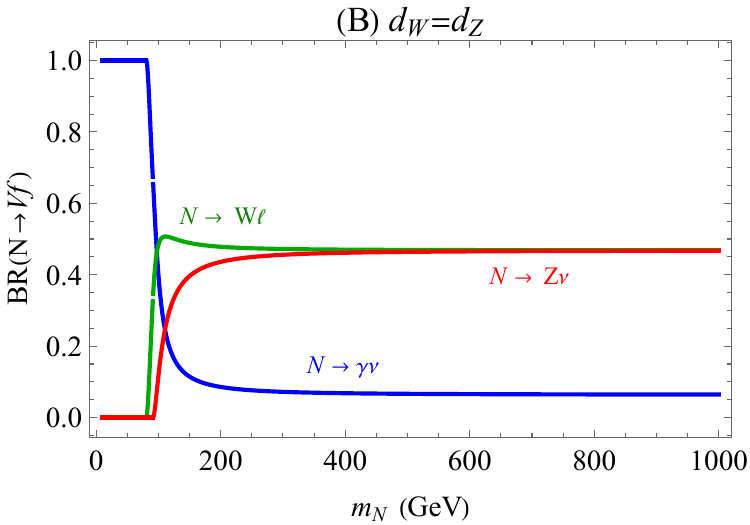}\\ \vspace{0.2cm}
 \includegraphics[scale=0.5]{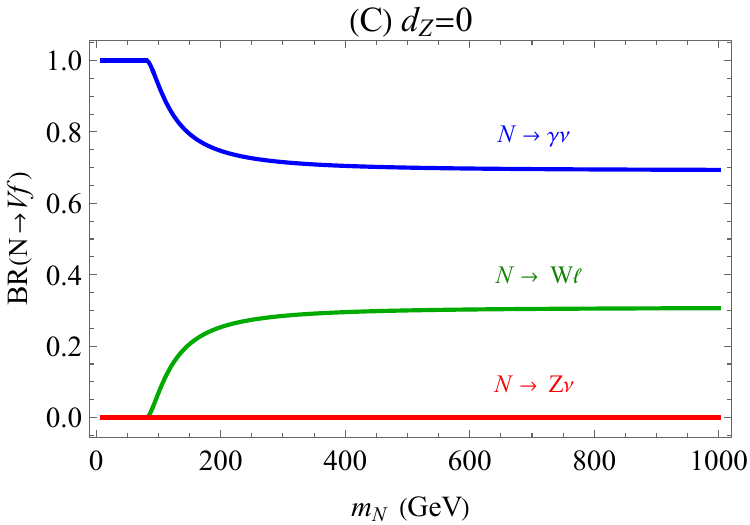} \hspace{0.2cm}\includegraphics[scale=0.5]{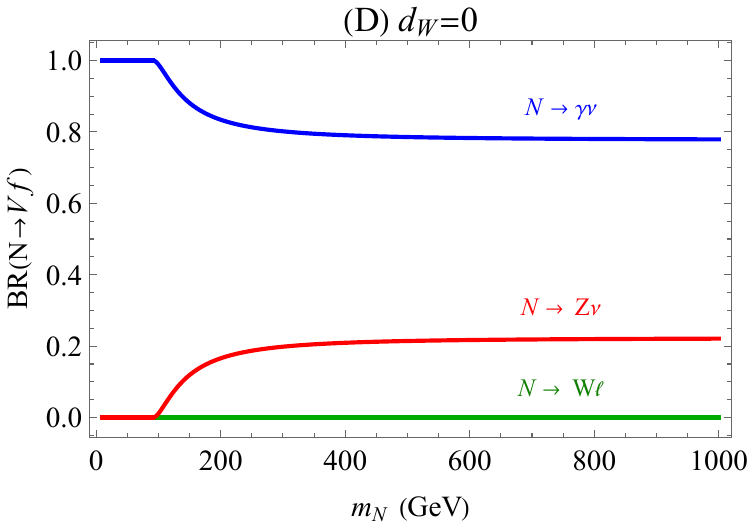}
 \caption{\em \small Decay branching ratios of the sterile neutrino, as function of its mass, for the four benchmarks: (A) $\mathcal{C}_B=0$, (B) $d_W=d_Z$, (C) $d_Z=0$ and (D) $d_W=0$.}\label{fig:BRs}
 \end{figure}

 We show in Fig~\ref{fig:BRs} the $N$ decay BRs for the different benchmarks and for a specific flavor $i$. We can see that $N \to \gamma \nu$ is the dominant decay below the threshold for decays into $W\ell$, $m_N < m_W$, and for scenarios with smaller $\mathcal{C}_W$'s. Above the $W$ mass, the decay $N \to W^\mp \ell^{i\, \pm}$ has a zero BR, as expected, in the scenario (D) $d_W=0$ but it becomes significant already in the scenario (C) $d_Z=0$, while it is the dominant decay in the scenarios with larger $\mathcal{C}_W$'s. It should be noted that the detection of this leptonic decay at colliders would provide important additional information compared to the other decays into a boson plus a neutrino. In fact, by identifying the flavor of the emitted lepton, it would be possible to distinguish the specific flavor of the dipole operator responsible for the decay of $N$. 
 For simplicity, in our analyses we will consider a flavour-universal scenario with $d^e=d^\mu=d^\tau$.

\section{Hadron colliders}\label{sec:hadron}

In this section, we perform an analysis of the projected sensitivities on the dipole-portal operator of future hadron colliders, the HL-LHC and the FCC-hh.

We rely on Monte Carlo simulations. Signal and background events are simulated with {\tt MadGraph5} \cite{Alwall:2014hca}. For the signal, the relevant interactions are implemented in {\tt MadGraph5} using {\tt Feynrules} \cite{Alloul:2013bka}. The events are then passed to {\tt Pythia8} \cite{Bierlich:2022pfr} for showering. Jets are clustered with {\tt Fastjet} \cite{Cacciari:2011ma} using an anti-kt algorithm with cone size $R$ = 0.5. To account for main detector effects, a smear is also applied to the 4-momenta of the jets, following the {\tt Delphes} \cite{deFavereau:2013fsa} default card. 
Moreover, when relevant, the missing transverse energy of each event is computed by including a Gaussian resolution $\sigma(E^{miss}_T)=0.49 \sqrt{\Sigma_i E^i_T}$, where $\Sigma_i E^i_T$ is the scalar sum of the transverse energies of all the reconstructed objects.

Statistical significance is assessed according to  $\sigma\equiv N_S /\sqrt{N_S+N_B}$, where $N_S$ ($N_B$) represents the number of signal (background) events.

\subsection{Sterile neutrino production at hadron colliders}
\label{sec:production-hadron}

\begin{figure}[h]
\centering
 \includegraphics[scale=0.6]{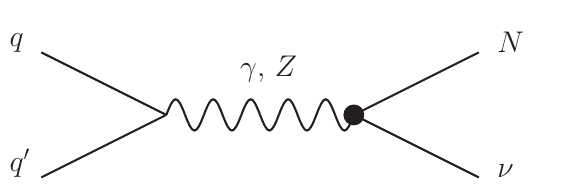} \includegraphics[scale=0.6]{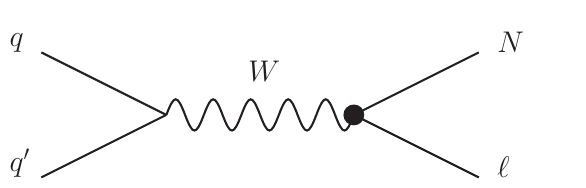}
 \caption{\em \small Feynman diagrams for the $N$ production via $s$-channels at hadron colliders. Left diagram: $pp\to N \nu$ induced by the $\bar{\nu}\sigma^{\mu\nu}N F_{\mu\nu}$ or the $\bar{\nu}\sigma^{\mu\nu}NZ_{\mu\nu}$ dipole operator, whose amplitude is proportional to the coupling $d_\gamma$ or $d_Z$ respectively. Right diagram: $pp \to N \ell$ induced by the $\bar{\ell}_L\sigma^{\mu\nu}N W_{\mu\nu}$ dipole operator and with amplitude proportional to $d_W$.}\label{fig:pp-s-channel}
 \end{figure}

\begin{figure}[h]
\centering
 \includegraphics[scale=0.55]{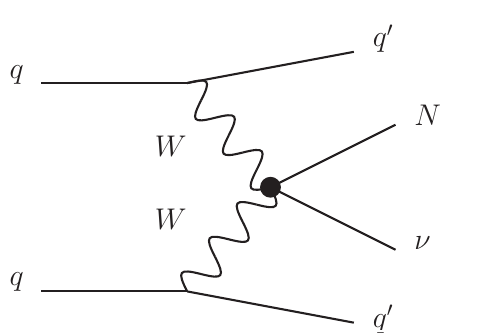} 
 \includegraphics[scale=0.55]{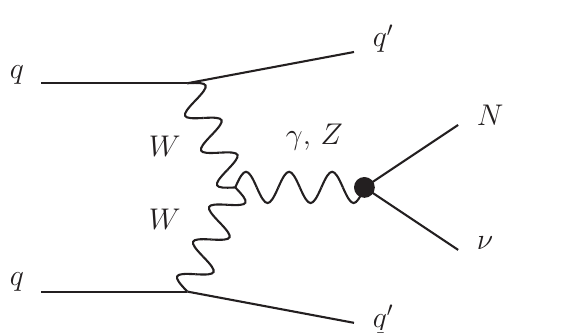}
 \includegraphics[scale=0.52]{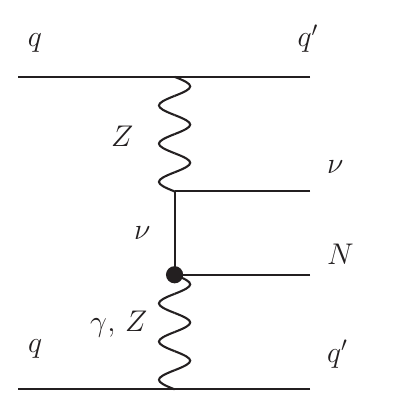} \\ \vspace{0.5cm}
 \includegraphics[scale=0.55]{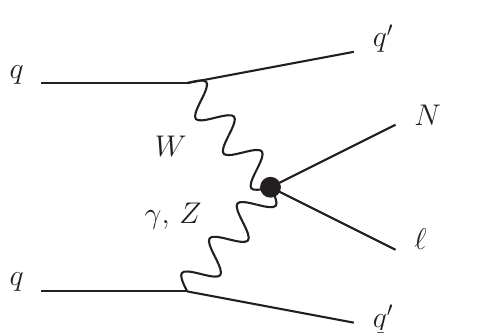}
 \includegraphics[scale=0.55]{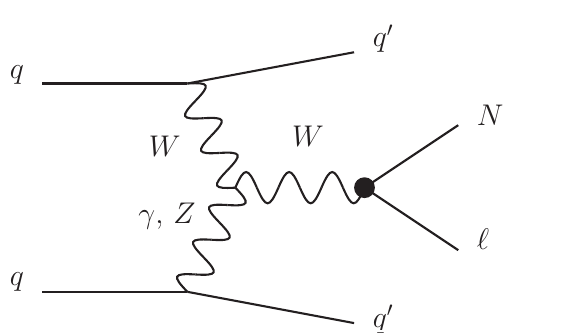}
 \includegraphics[scale=0.52]{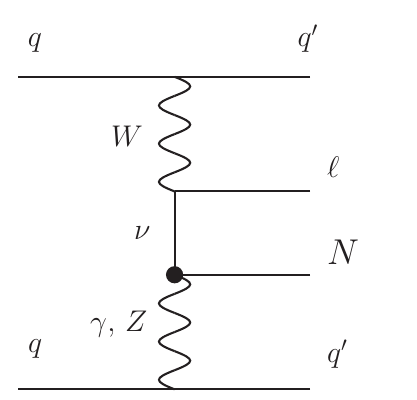}
 \caption{\em \small  
 Representative Feynman diagrams for the production of $N$ via VBF at hadron colliders. Upper diagrams: VBF $N$ production in association with a neutrino. Lower diagrams: VBF $N$ production in association with a lepton. The leftmost diagrams in the figure are generated by two-bosons effective interactions induced by the expansion of the dipole operator $\mathcal{O}_W$ (second line in Eq.~\eqref{eq:dipoles}) and have amplitudes proportional to $d_W$. The other diagrams are controlled by single-boson effective interactions induced by $\bar{\nu}_L\sigma^{\mu\nu}N F_{\mu\nu}$, $\bar{\nu}_L\sigma^{\mu\nu}N Z_{\mu\nu}$ or $\bar{\ell}_L\sigma^{\mu\nu}N W_{\mu\nu}$ and with amplitudes proportional to $d_\gamma$, $d_Z$ or $d_W$ respectively.
 }\label{fig:pp-vbf}
 \end{figure}

Sterile neutrino production at hadron colliders can occur through two main mechanisms. The first category, as shown in Fig.~\ref{fig:pp-s-channel}, is production through $s$-channel exchange of neutral gauge bosons $\gamma,Z$, leading to processes $pp \to N\nu$, or via the charged boson $W$, leading to the $N$ associated production with a lepton, $pp \to N \ell$. These processes are mediated by dipole operators with single-boson effective interactions 
(first line in Eq.~\eqref{eq:dipoles}) and have amplitudes proportional to the couplings $d_\gamma$, $d_Z$ and $d_W$, respectively.
The second category of processes, which, as we will see, is particularly relevant at higher energies, consists of the production of $N$ by VBF, shown in Fig.~\ref{fig:pp-vbf}.  Both the production of $N$ accompanied by a neutrino and that accompanied by a lepton are possible. 
 These VBF processes result from the interplay of several different contributions, where the $N$ production can be mediated by single-boson effective interactions, induced by the operators $\bar{\nu}_L\sigma^{\mu\nu}N F_{\mu\nu}$, $\bar{\nu}_L\sigma^{\mu\nu}N Z_{\mu\nu}$ or $\bar{\ell}_L\sigma^{\mu\nu}N W_{\mu\nu}$, or by two-bosons effective interactions induced by the operator $\mathcal{O}_W$ (second line in Eq.~\eqref{eq:dipoles}). 
 It is clear that to correctly evaluate $N$ production via VBF, preserving gauge invariance, it is necessary to include all possible contributions. In particular, the effective interactions with two gauge bosons resulting from the expansion of the operator $\mathcal{O}_W$  cannot be neglected.
   
We remark some aspects of the $N$ production at hadron colliders. $N$ production can occur for any flavor of the dipole operator. As already specified, in the following analysis we will consider a universal flavor scenario.
In the case of the $N$ associated production with a lepton, the flavor of this lepton will be the same as that of the operator. By tagging its flavor it is therefore possible to extract information on the flavor of the dipole operator that allows the production of $N$. For simplicity, we will focus on processes with only electrons or muons, neglecting taus. Although, once the tau decays are properly analyzed, we expect similar sensitivities for the tau-flavored dipole operator.

\subsection{The High-Luminosity LHC}\label{sec:hl-lhc}

\begin{figure}[h!]
\centering
 \includegraphics[scale=0.5]{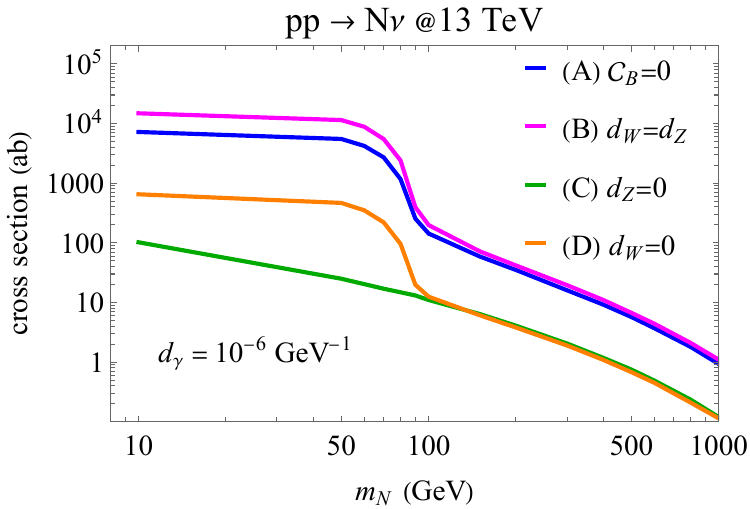} \includegraphics[scale=0.5]{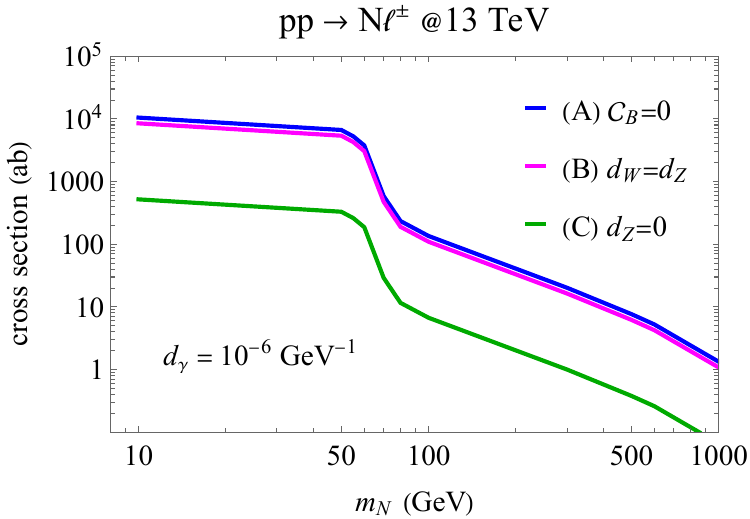}
 \caption{\small \em Cross section for the $N$ production associated with a neutrino (left) and with a lepton, either a muon or an electron (right) at the HL-LHC for the different theoretical scenarios (A)-(D). Cross sections are shown for a reference value $d_\gamma=10^{-6}$ GeV$^{-1}$. 
  }\label{fig:xsec-13}
 \end{figure}

We start by evaluating the future sensitivities on $d_\gamma$ of the HL-LHC. Conservatively, we consider a collision energy of 13 TeV and a final collected integrated luminosity of 3 ab$^{-1}$. The main $N$ production at the HL-LHC is through $s$ channels. $N$ production through VBF would be possible in principle, but is subleading and negligible at the HL-LHC collision energy.  We show in Fig.~\ref{fig:xsec-13} the cross section for the $N$ production at the HL-LHC in the different theoretical benchmarks.

 Processes with subsequent decays $N\to \gamma \nu$ have already been analyzed in other studies \cite{Magill:2018jla, Brdar:2025iua}, which find sensitivities of the order of $d_\gamma \sim 3 \times 10^{-5}$ GeV$^{-1}$ in scenario (D) $d_W=0$. In our analysis, we instead want to consider the decay $N\to W\ell$ in the remaining scenarios with $d_W \neq 0$. As already highlighted, this channel has the advantage of allowing the detection of the flavor related to the dipole operator responsible for the decay of $N$.
We will consider in particular the production by $s$-channel $W$ exchange, which combined with the leptonic decay of $N$ leads to the following possible final states:
\begin{equation}
pp \to N \ell^{i\, \pm} \to W^\pm \ell^{j \, \mp}\ell^{i\, \pm}\; , \, W^\mp \ell^{j \, \pm}\ell^{i\, \pm} \, \text{(SSD)}
\end{equation}
we will then focus on events with hadronically decaying $W$'s and with $\ell \equiv e,\mu$.\\
Notice that if the flavor of the $N$ production operator is different from the one in the $N$ decay ($i\neq j$), we will have two leptons with different flavors in the final state. Moreover, the two leptons will be kinematically different and thus distinguishable. Also note that, due to the Majorana nature of $N$, half of the decays will occur in two leptons with the same charge.
Therefore, the process we will analyze could allow us to obtain multiple information: the chance of detecting and distinguishing the flavors of possibly different dipole operators responsible for the production and decay of $N$, the possibility of establishing the Majorana nature of the heavy neutrino.

We then focus on the final state with same-sign dileptons $ W_h^\mp \ell^{ \pm}\ell^{ \pm}$ (SSD), which enjoys a rather low background, mainly given by $Z_\ell (\gamma^\star_\ell)$+jets and $Z_\ell (\gamma^\star_\ell) V_h$ events ($V_h\equiv W_h,Z_h$), where the charge of one of the leptons from $Z(\gamma^\star)$ decays is misidentified.\footnote{We consider a misidentification rate of 0.1\%~\cite{ATLAS:2019jvq}.} Additional 
background components are given by events with two gauge bosons decaying into leptons accompanied by jets, $W^\pm_\ell W^\pm_\ell (W^\pm_\ell Z_\ell)$+jets, and by the production of three gauge bosons, two of which decay leptonically and the other hadronically, $W^\pm_\ell W^\pm_\ell (W^\pm_\ell Z_\ell)V_h$. 
We require exactly two leptons with the same charge and at least two jets in the final state with transverse momenta $p_T>20$ GeV and emitted in the central region, $|\eta|<2.5$. We also require a minimal separation $\Delta R(\ell, j)>0.4$ between the tagged jets and the leptons. 
For the signal, the two leading-$p_T$ jets correctly reconstruct $W_h$. 
To reduce the background components $Z_\ell(\gamma^\star_\ell)$+jets and $W_\ell W_\ell (W_\ell Z_\ell)$+jets we apply the following condition on the invariant mass of the two leading-$p_T$ jets: $M_{jj} \in $~[60 GeV, 90 GeV]. To reduce background sources from lepton charge misidentification, in particular $Z_\ell$+jets, we apply a cut on the invariant mass of the two leptons: $M_{\ell^{(1)}\ell^{(2)}}>100$ GeV. After these acceptance criteria, the background is overall at the level of 7.9 fb.

\begin{figure}[h!]
\centering
 \includegraphics[scale=0.37]{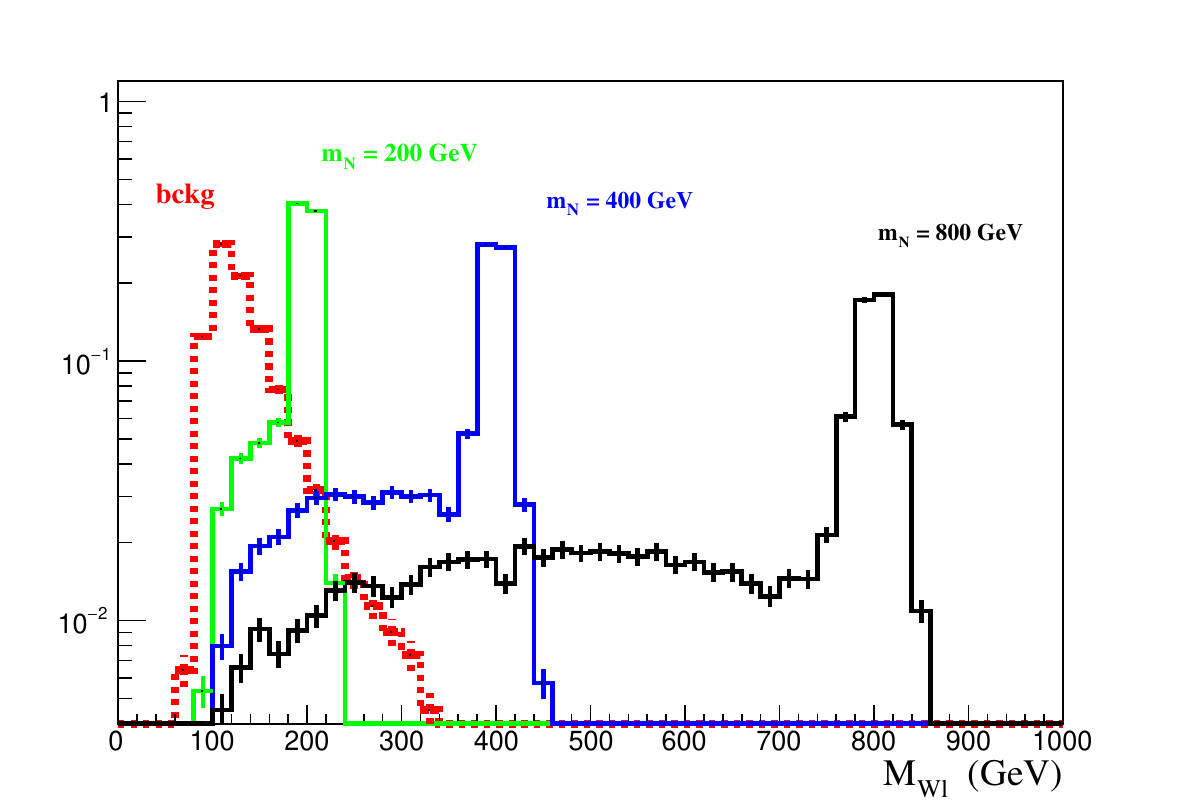} \includegraphics[scale=0.35]{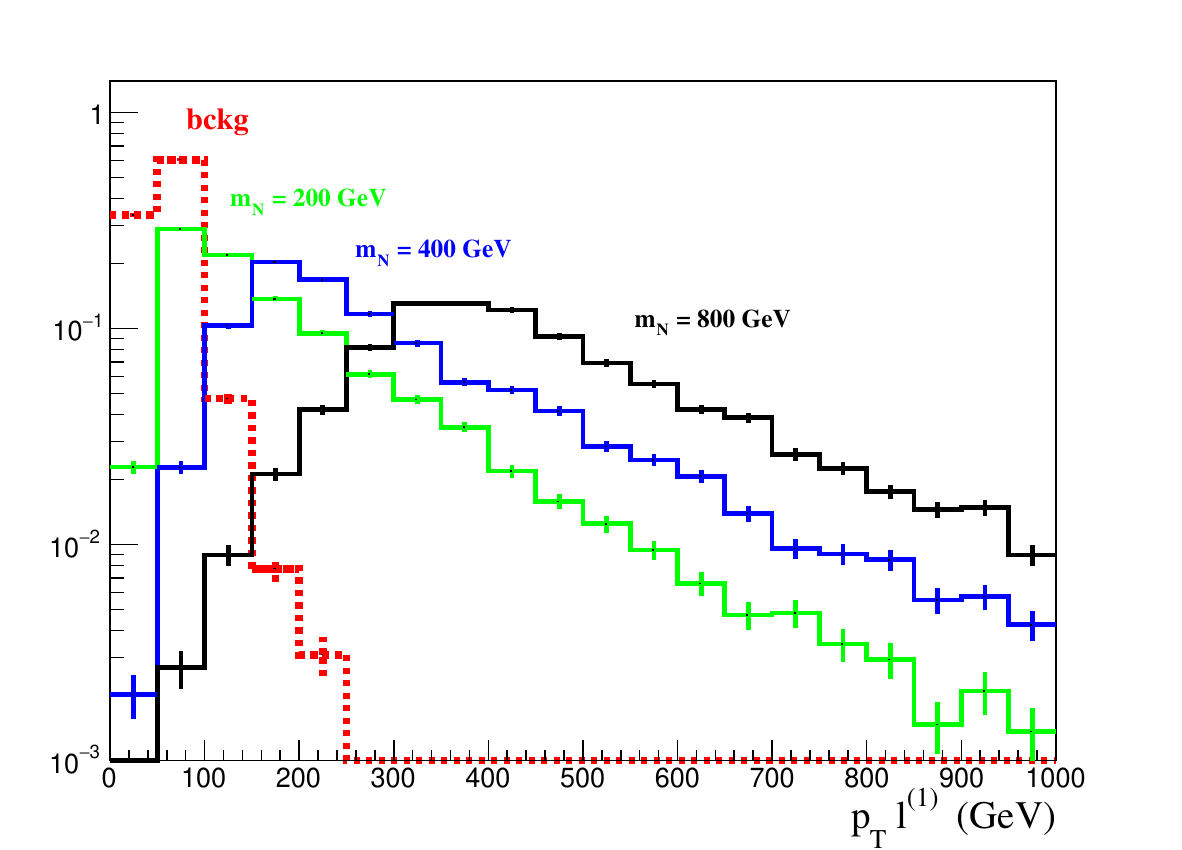}
 \caption{\small \em Distributions normalized to unit area for the $pp\to W^\mp_h\ell^\pm\ell^\pm$ neutrino dipole-portal signal with different $N$ masses and for the SM background.  Left plot: invariant mass of the system consisting of the hadronically decaying $W$ and the nearest (in $\Delta R$) lepton. Right plot: transverse momentum distribution of the $p_T$-leading lepton.}\label{fig:dist-hllhc}
 \end{figure}

We apply the following simple strategy to further reduce the background and identify the signal. The lepton resulting from the decay of $N$ is for the vast majority of signal events the one closest to $W_h$, that is, the one with the smallest separation $\Delta R=\sqrt{\Delta\phi^2+\Delta\eta^2}$ from $W_h$. In this way we can distinguish the lepton emitted in the decay of $N$ from the one produced with it in association, and we can reconstruct quite well, as shown in figure \ref{fig:dist-hllhc}, the sterile neutrino, starting from the invariant mass of its decay products: the $W_h$ and the lepton just identified, $M_{W_h \ell}$. Moreover, for the range of $N$ masses we are considering, the decay products are generally more energetic than for the background (Fig.~\ref{fig:dist-hllhc}). We can then further attenuate the background by applying a cut in the transverse momentum of the leading-$p_T$ lepton. 
We then apply, depending on the mass of $N$, the three sets of cuts shown in Table \ref{tab:cut-hllhc}. We choose these cuts such that the signal with the lightest $N$, for each set, has an efficiency of at least 80\%.

\begin{table}[h!]
\centering
\begin{tabular}{|c|c|c|c|}
\hline
$m_N$ in GeV & [150, 300) & [300, 500) & $\ge$ 500 \\
\hline
cuts in GeV & $p_T \ell^{(1)}>$ 50 , $M_{W_h \ell}>$ 100 & $p_T \ell^{(1)}>$ 100 , $M_{W_h \ell}>$ 200 & $p_T \ell^{(1)}>$ 150 , $M_{W_h \ell}>$ 400 \\
\hline
Bckg cross sec. & 5.7 fb & 0.65 fb & 0.097 fb \\ 
\hline
\end{tabular}
 \caption{\small \em Set of cuts applied in signal selection for the different range of $N$ masses at the HL-LHC. The background cross section after cuts is also reported. }\label{tab:cut-hllhc}
\end{table}

\begin{figure}[h!]
\centering
 \includegraphics[scale=0.7]{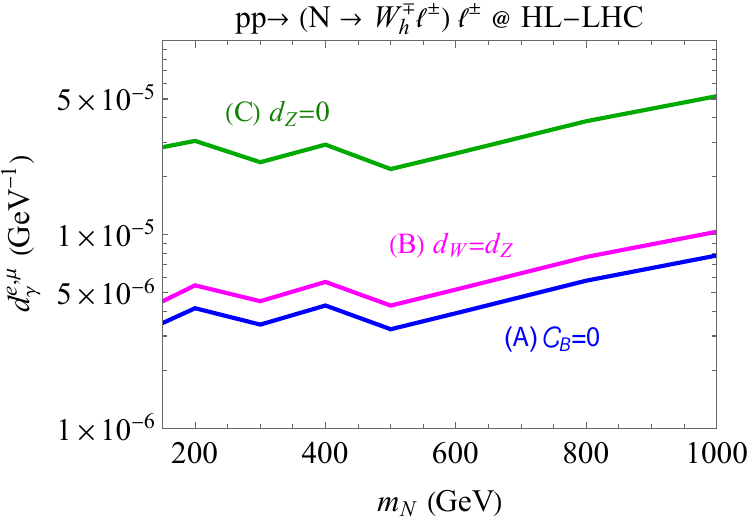} 
 \caption{\small \em HL-LHC 2-$\sigma$ sensitivities to the sterile neutrino dipole portal operator in the channel  $pp\to W^\mp_h \ell^\pm \ell^\pm$
 for the different theoretical benchmarks (A)-(C). A flavor-universal scenario $d^e_\gamma=d^\mu_\gamma=d^\tau_\gamma$ is assumed.}\label{fig:sensitivity-hllhc}
 \end{figure}

After the final selection, we find the 2$\sigma$ sensitivities to the neutrino dipole portal operator shown in Figure \ref{fig:sensitivity-hllhc}. We recall that since we focused on electrons or muons in the final state in a flavor-universal scenario, our analysis is equally sensitive to muonic and electronic flavors, $d^{e,\mu}_\gamma$. We can observe that the sensitivities in scenario (C) $d_Z=0$ are similar to those estimated in the literature for scenario (D) $d_W=0$, considering the decay channel $N\to \gamma\nu$. In the remaining benchmarks (B) $d_W=d_Z$ and (A) $\mathcal{C}_B=0$ the sensitivities are significantly better, up to one order of magnitude, reaching values 
$d_\gamma \sim 3\times 10^{-6}$ GeV$^{-1}$.

\subsection{The FCC-hh}\label{sec:fcc}

We investigate now the sensitivity to the neutrino dipole portal of the future circular collider FCC-hh, with a nominal collision energy of 100 TeV. We assume an integrated collected luminosity of 30 ab$^{-1}$, based on the expected machine performances \cite{FCC:2018vvp}. 

The scenario for sterile neutrino production changes significantly with the increasing of energy from the HL-LHC case, and we expect, in particular, an enhancement of the VBF production mechanisms. This is a general feature of future collider experiments at the energy frontier such as FCC-hh or the muon collider, which has been already emphasized in the literature (see for example \cite{Mohan:2015doa, Goncalves:2017gzy, Baker:2022zxv, Molinaro:2017mwb, Costantini:2020stv}).

\begin{figure}[h]
\centering
 \includegraphics[scale=0.5]{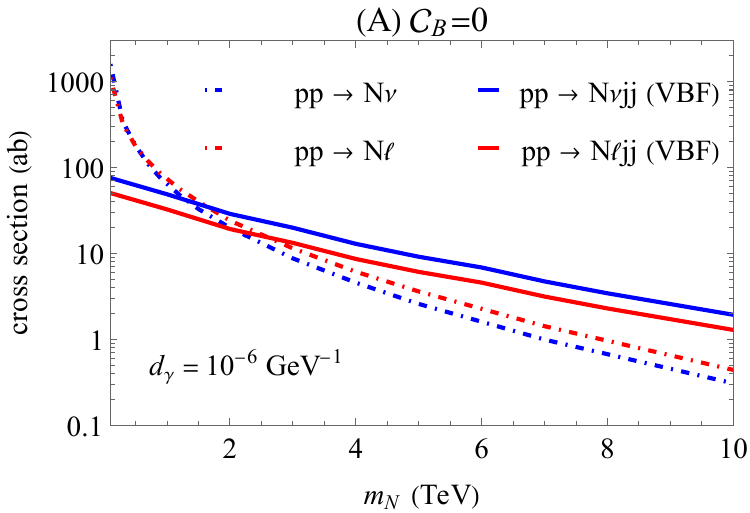} \includegraphics[scale=0.5]{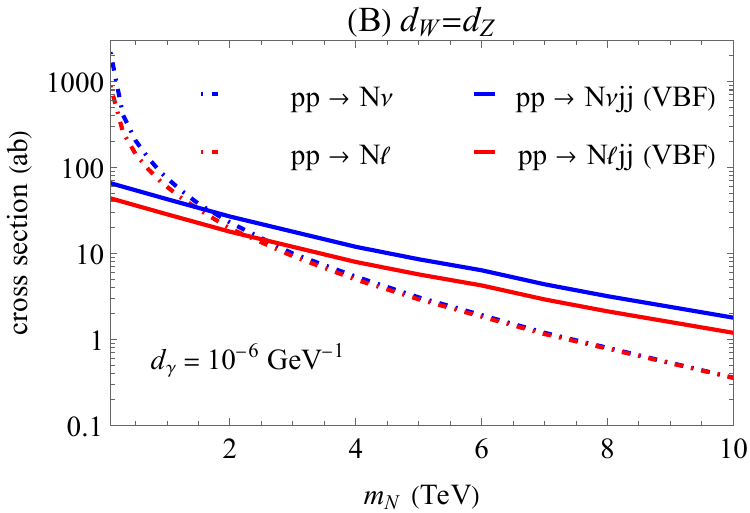}\\ \vspace{0.2cm}
 \includegraphics[scale=0.5]{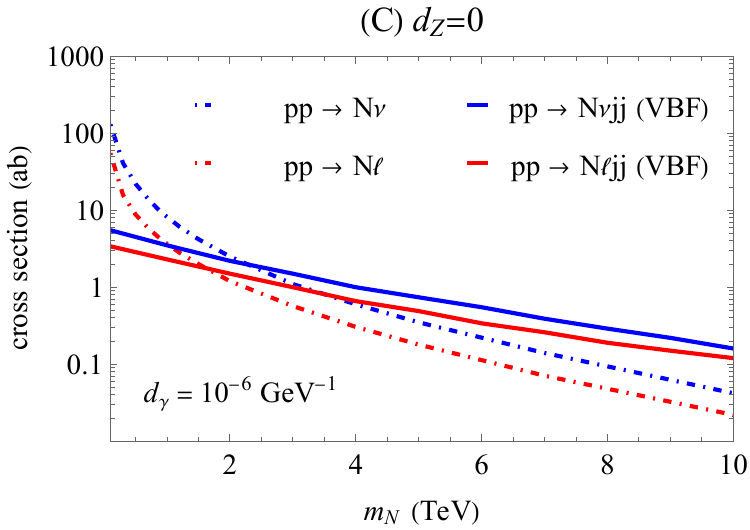} \includegraphics[scale=0.5]{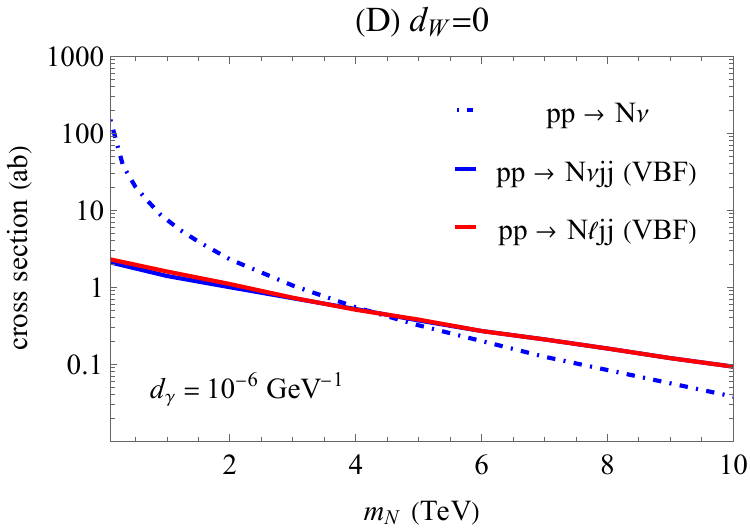}
 \caption{\small \em Cross section for the $N$ production at the FCC-hh in the different theoretical benchmarks and channels: associated production with a neutrino (blue curves) or with a lepton (red curves) via $s$ channels (dot-dashed curves) or through VBF (continuous curves). Cross sections are shown for a reference value $d_\gamma=10^{-6}$ GeV$^{-1}$.  }\label{fig:xsec-fcc}
 \end{figure}

We show in Fig.~\ref{fig:xsec-fcc} cross section values for the production of $N$ in the different channels, the associated production with a neutrino or with a lepton through $s$ channels or VBF, and for the theoretical benchmarks (A)-(D). We apply minimal requirements on the lepton in the associated production $\ell N$: $p_T>20$ GeV, $|\eta|<2.5$ and on the jets emitted in VBF processes: $p_T>20$ GeV, $|\eta|<5$.
We can observe that VBF becomes significant for $m_N\gtrsim 1$ TeV and is the leading production mechanism for the heavier sterile neutrinos. VBF is more significant for the theoretical scenarios with larger values of $\mathcal{C}_W$. This is expected due to the dominant role that the dipole operator $\mathcal{O}_W$ plays in VBF processes. For lighter $N$'s, $s$-channel production is dominant, but the corresponding cross sections undergo a rapid decline as $m_N$ increases. 
Given these considerations, we will analyze $s$ channels, and, for the heavier $N$'s, with masses greater than 1 TeV, we will also study the VBF processes. 

\paragraph{$s$-channel \\}
For the investigation in the $s$ channels, in particular, we retrace the analysis performed for the HL-LHC in the {\bf SSD} channel $pp \to \mathbf{W_h^\pm\ell^\mp\ell^\mp}$. The main difference is a significant increase in the boost of the final products at FCC-hh. $W_h$ thus 
decays into a single fat-jet for the vast majority (more than 80\%) of signal events with $m_N\geq 1$ TeV. 
For these heavier $N$'s we therefore identify the $W_h$ of the signal with the leading-$p_T$ jet in the final state 
and we require its invariant mass to lie in the interval [60 GeV, 90 GeV].\footnote{This requirement strongly reduces in particular the $Z_\ell(\gamma^\star_\ell)$+jets background component, for which the leading-$p_T$ jet has generally an invariant mass lower than 60 GeV.} For the signal with $m_N<1$ TeV the $W_h$ is reconstructed from the leading- and the second leading-$p_T$ jets.  To attenuate background sources from misidentified lepton charges, we also apply the condition $M_{\ell^{(1)}\ell^{(2)}}>100$ GeV.
The sterile neutrino can be efficiently reconstructed with the same strategy outlined in the previous section for HL-LHC, by considering the $W_h$ and the nearest lepton. We also apply the same set of cuts for the lighter $N$'s. 
Note that this is a conservative choice because, again due to the greater boost, the leading-$p_T$ lepton of the signal is more energetic (generally 20\% more) than in the case of HL-LHC and therefore even more stringent cuts could be applied. The background, again dominantly given by $Z_\ell(\gamma^\star_\ell)$+jets and $Z_\ell(\gamma^\star_\ell) V_h$ ($V_h\equiv W_h, Z_h$) events with misidentified lepton charges and by $W^\pm_\ell W^\pm_\ell (W^\pm_\ell Z_\ell)$+jets and $W^\pm_\ell W^\pm_\ell (W^\pm_\ell Z_\ell)V_h$, has a total cross section of 102 fb after acceptance requirements. The total background cross section reduces as shown in Tab.~\ref{tab:cut-fcc} after each set of cuts. Additionally, for $N$ heavier than 1 TeV we will apply the cut: $M_{W_h \ell}>90\%\cdot m_N$.

\begin{table}[h!]
\centering
\begin{tabular}{|c|c|c|c|c|}
\hline
$m_N$ in GeV & [150, 300) & [300, 500) & [500, 1000) & 1000 \\
\hline
cuts in GeV ($p_T \ell^{(1)}$, $M_{W_h \ell}$)&  $>$50 ,  $>$100 & $>$100 , $>$200 & $>$150 , $>$400 & $>$200 , $>$900 \\
\hline
Bckg cross sec. & 78 fb & 13 fb & 2.2 fb & 0.092 fb\\ 
\hline
\end{tabular}
 \caption{\small \em Set of cuts applied in signal selection in the $s$-channel $W^\pm_h\ell^\mp\ell^\mp$ for the different range of $N$ masses at the FCC-hh. The background cross section after cuts is also reported. }\label{tab:cut-fcc}
\end{table}

In scenario (D) $d_W=0$ the SSD channel sofar considered is absent, but we can derive a rough estimate of the FCC-hh sensitivity starting from the HL-LHC estimate made in the literature considering the decay channel $N\to \nu\gamma$  \cite{Magill:2018jla,Brdar:2025iua}. Based on a simple scaling of the signal and the dominant background cross sections, and taking into account the different integrated luminosities, we find sensitivities in the range of $d_\gamma \sim 6 \times10^{-6}$ GeV$^{-1}$ for the FCC-hh in the scenario $d_W=0$ and $m_N\lesssim 1$ TeV. \\

We will now outline signal selection strategies for {\bf VBF processes}. 
Several VBF channels can be analyzed. Depending on $N$ production and decay modes, we can list the following possible final states:
\begin{multicols}{2}
    \begin{enumerate}[label=(\roman*)]
        \item $(N\to W^\pm \ell^\mp)\nu jj \to  W^\pm \ell^\mp jj +E^{miss} $
           \item $(N\to W \ell)\ell jj \to W \ell \ell jj $ (SSD+OSD)
         \item $(N\to \gamma\nu)\ell^\pm jj \to \gamma \ell^\pm jj +E^{miss} $
          \item $(N\to \gamma\nu)\nu jj \to \gamma jj +E^{miss} $
           \item $(N\to Z\nu)\ell^\pm jj \to Z \ell^\pm jj +E^{miss} $
          \item $(N\to Z\nu)\nu jj \to Z jj +E^{miss} $
    \end{enumerate}
    \end{multicols}
    
where $E^{miss}$ denotes the missing energy associated with neutrinos. Note that channel (ii) includes final states with dileptons of both opposite sign charges (OSD) and same sign (SSD). The latter arise from the Majorana nature of $N$.

An intertwined analysis of all these final states would be particularly interesting to extract multiple information and increase the sensitivity to the dipole portal operator. As already highlighted, channels (ii) and (iii) where $N$ is produced in association with a lepton could allow, for example, the detection of the flavor of the dipole operator which induces $N$ production. For effectiveness and concreteness, we focus in this study on some of the most interesting channels and defer a joint analysis to future investigations. 
Given the higher decay BRs and the possibility of reconstructing the sterile neutrino from its visible decay products, we will focus on channels (i) and (ii), with final hadronically decaying $W$'s, for the theoretical scenarios (A) $\mathcal{C}_B=0$ and (B) $d_W=d_Z$. In the analysis of channel (ii), we will focus in particular on the SSD final state, which enjoys a lower SM background. 
Due to the reduced or absent decay BRs in $W\ell$, for scenarios (C) $d_Z=0$ and (D) $d_W=0$ we will instead focus on the final state (iii), where the sterile neutrino decays into a neutrino plus a photon.

\begin{figure}[h]
\centering
 \includegraphics[scale=0.35]{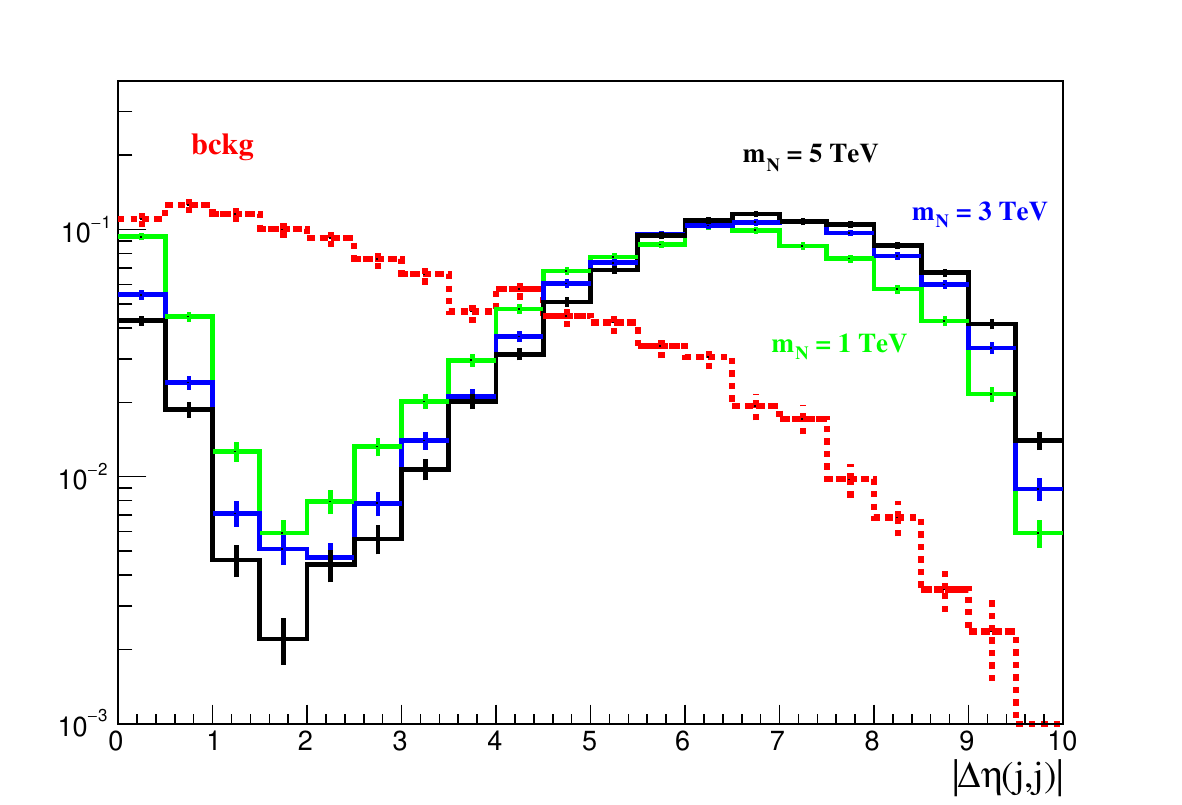} \includegraphics[scale=0.35]{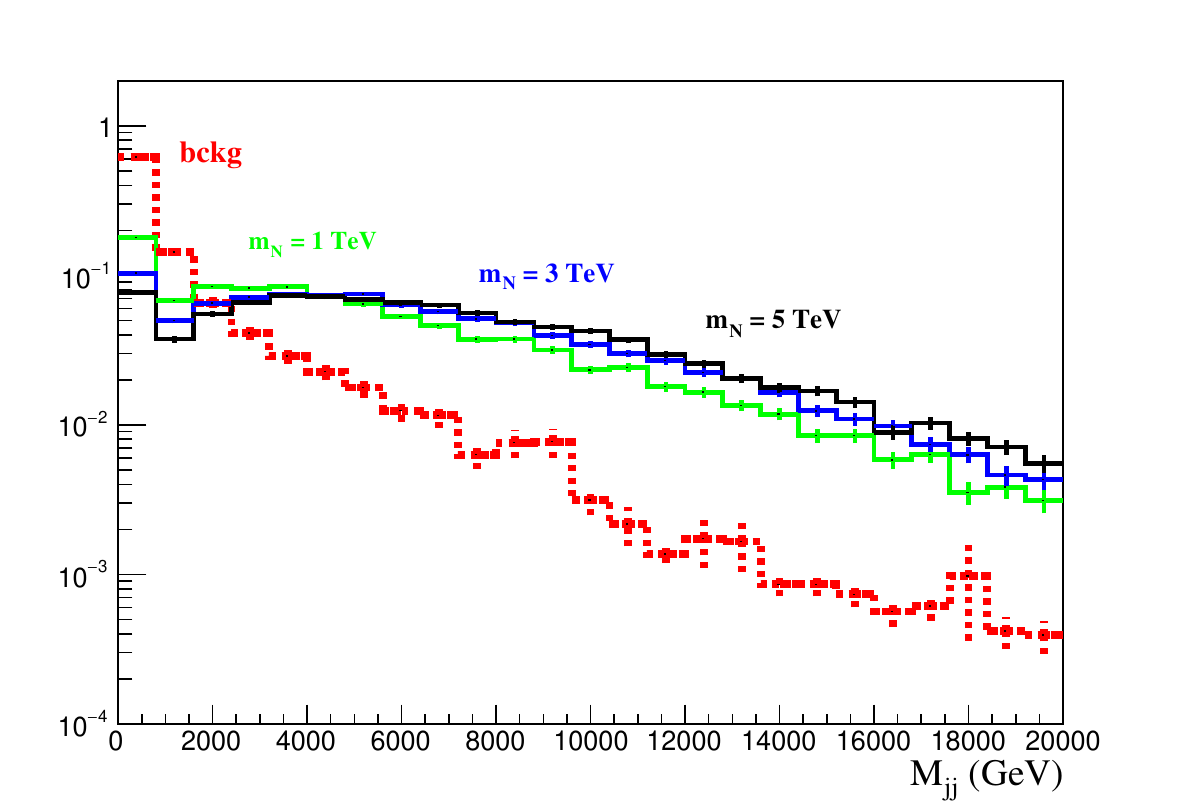}
 \caption{\small \em Distributions normalized to unit area for the $pp\to W_h^\pm \ell^\mp \ell^\mp jj$ VBF dipole-portal signal in scenario (A) $\mathcal{C}_B=0$ with different $N$ masses, and for the SM background. We plot the pseudorapidity separation of the two VBF jets (left plot) and their invariant mass (right plot).
 }\label{fig:VBF-dist}
 \end{figure}

A common feature of these VBF channels, which we exploit to select the sterile neutrino signal from the background, is the peculiar kinematics of the two final jets. As a characteristic of VBF processes, these jets are emitted forward-backward at high rapidity and therefore have a separation $\Delta\eta$ generally much higher than for the background. The invariant mass of the two jets tends to be higher as well. We show in Fig.~\ref{fig:VBF-dist} the distributions $\Delta\eta(j,j)$ and $M_{jj}$ in the illustrative case of signal (ii) for the theoretical scenario (A) $\mathcal{C}_B=0$. Very similar distributions are obtained for the other VBF channels and theoretical benchmarks.

In all the VBF channels we will analyze, we will apply the following cuts on the VBF jets:
\begin{equation}\label{eq:VBF-cut}
    |\Delta \eta(j, j)|>3 \;,\quad  M_{jj}>500 \; \text{GeV}
\end{equation}

Another common feature is the fact that at FCC-hh, as already underlined, $W_h$ decays into a single, central, high-$p_T$ fat jet for the vast majority (more than 80\%) of signal events with $m_N\geq 1$ TeV. The two jets emitted by VBF are instead softer (as well as generally emitted at higher rapidity). In the following analyses, $W_h$ will therefore be identified with the leading-$p_T$ jet of the final state, and we will require that its invariant mass is in the range 60-90 GeV.
Other selection strategies are discussed channel by channel.

\paragraph{(i) $\bf{W_h^\pm \ell^\mp jj +E^{miss}}$ \\}
We require exactly one lepton with $p_T>20$ GeV and $|\eta|<2.5$ and at least three jets with $p_T>20$ GeV, $|\eta|<5$ and separated from each other and from the lepton by $\Delta R(j,j),\Delta R(j,\ell)>0.4$. 
The dominant background is given by $V_h W_\ell$+jets events, where the jets (at least two) are produced by QCD, or by EW interactions. This latter category includes the EW production of two gauge bosons, via VBF, and of three gauge bosons. An additional background component arises from $W_\ell$+jets, with at least three QCD jets. The latter process has a cross section of about 10$^4$ pb after minimal acceptance cuts, which is, however, strongly reduced to the level of 10 pb, by imposing that the leading-$p_T$ jet has an invariant mass between 60 and 90 GeV, so as to mimic the $W_h$ of the signal. The cross sections, before selection, of the remaining $V_h W_\ell$+jets backgrounds are about 130 pb for the events with QCD jets and about 4.4 pb for the component with EW jets.  
As a first step of signal selection, we apply the VBF cuts in Eq.~\eqref{eq:VBF-cut}. The total background cross section is then reduced to 23 pb, while more than 80\% of the signal events are retained. In the second step, the sterile neutrino can be easily reconstructed, from the $W_h$ and the lepton, and we can exploit the characteristic of the signal to emit energetic final states. We apply the cuts:
\[
p_T \ell>200 \, \text{GeV}, \; E^{miss}_T>200 \, \text{GeV}, \; p_T W_h>200 \, \text{GeV}, \; M_{W_h\ell}>95\%\cdot m_N
\]
 The signal efficiency to these cuts (after passing the VBF selection) is greater than 90\% while the background cross section is reduced to less than 61 fb. 

\begin{figure}[h]
\centering
 \includegraphics[scale=0.35]{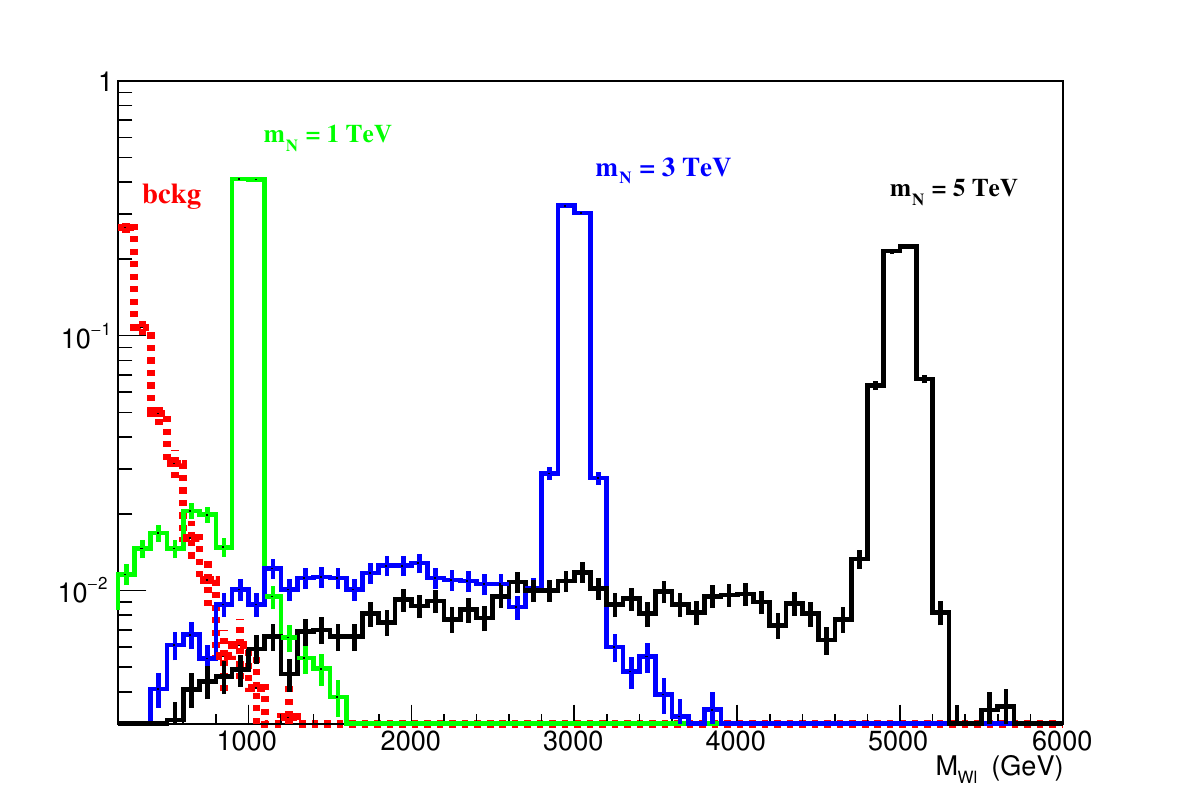} \includegraphics[scale=0.35]{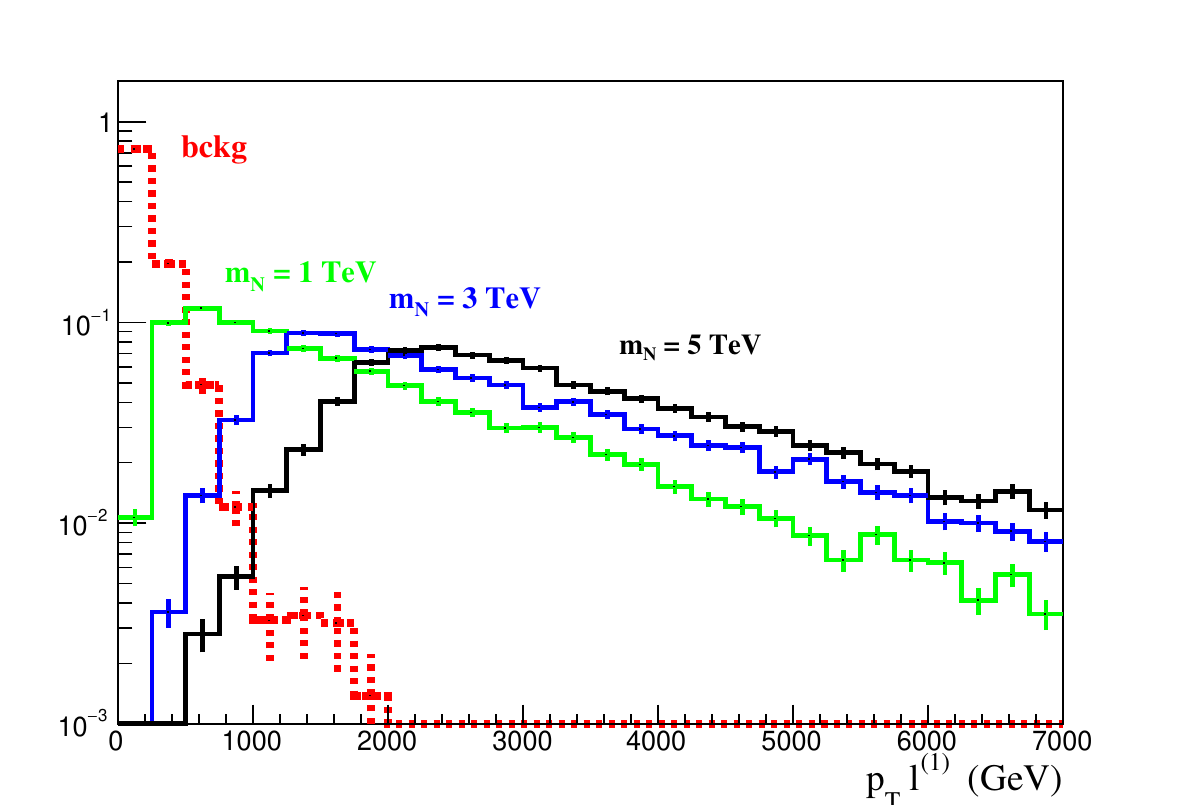}
 \caption{\small \em Distributions normalized to unit area for the $pp\to W_h^\pm \ell^\mp \ell^\mp jj$ VBF dipole-portal signal in scenario (A) $\mathcal{C}_B=0$ with different $N$ masses, and for the SM background. We plot the invariant mass of the system consisting of the $W_h$ and the nearer lepton (left plot) and the transverse momentum of the leading-$p_T$ lepton (right plot).
 }\label{fig:dist-SSDvbf}
 \end{figure}

\paragraph{(ii) $\bf{W_h^\pm \ell^\mp \ell^\mp jj}$ (SSD) \\ \noindent} 
Given the low background, consisting of only reducible components, we analyze the SSD channel.  We focus on a final state consisting of exactly two leptons with the same charge, and with $p_T>20$ GeV and $|\eta|<2.5$, and of at least three jets with $p_T>20$ GeV, $|\eta|<5$, and separated from each other and from the leptons by $\Delta R(j,j),\Delta R(j,\ell)>0.4$. The leading-$p_T$ jet, which is the $W_h$ of the signal, must have an invariant mass in the range 60-90 GeV.
Dominant background processes include $ W^\pm_\ell W^\pm_\ell V_h$+jets, where the jets, at least two, are produced by QCD or EW interactions and $Z_\ell(\gamma^\star_\ell)$+($\geq 3$) jets or $ Z_\ell (\gamma^\star_\ell) V_h$+($\geq 2$) jets events with misidentified lepton charges. The latter 
are strongly reduced by imposing the condition on the invariant mass of the two leptons: $M_{\ell^{(1)}\ell^{(2)}}>100$ GeV. $W^\pm_\ell W^\pm_\ell+(\geq 3)$jets is attenuated to a negligible level by the leading-$p_T$ invariant mass condition. After acceptance requirements, the background cross section is overall 50 fb. Similarly to the case of the SSD signal in the $s$ channel, the sterile neutrino can be efficiently reconstructed by considering the $W_h$ and the nearest (with the smaller $\Delta R$ separation) lepton. The corresponding invariant mass distributions, $M_{W_h\ell}$, are shown in 
Fig.~\ref{fig:dist-SSDvbf} for the signal with different $N$ masses and for the background. We plot signal distributions for scenario (A). Very similar distributions are obtained in the theoretical benchmark (B).

In our analysis, we first apply the VBF cuts in Eq.~\eqref{eq:VBF-cut} and then refine the selection by applying the following conditions on the leading-$p_T$ lepton and invariant mass of the reconstructed $N$:
\[
p_T\ell^{(1)}>200\,\text{GeV}, \;  M_{W_h\ell}>90\%\cdot m_N
\]
$p_T\ell^{(1)}$ distributions for signal and background are also shown in Fig.~\ref{fig:dist-SSDvbf}. Signal efficiencies for this selection are above 75\%, while the background cross section is reduced to less than 0.7 fb.

\begin{figure}[h]
\centering
 \includegraphics[scale=0.35]{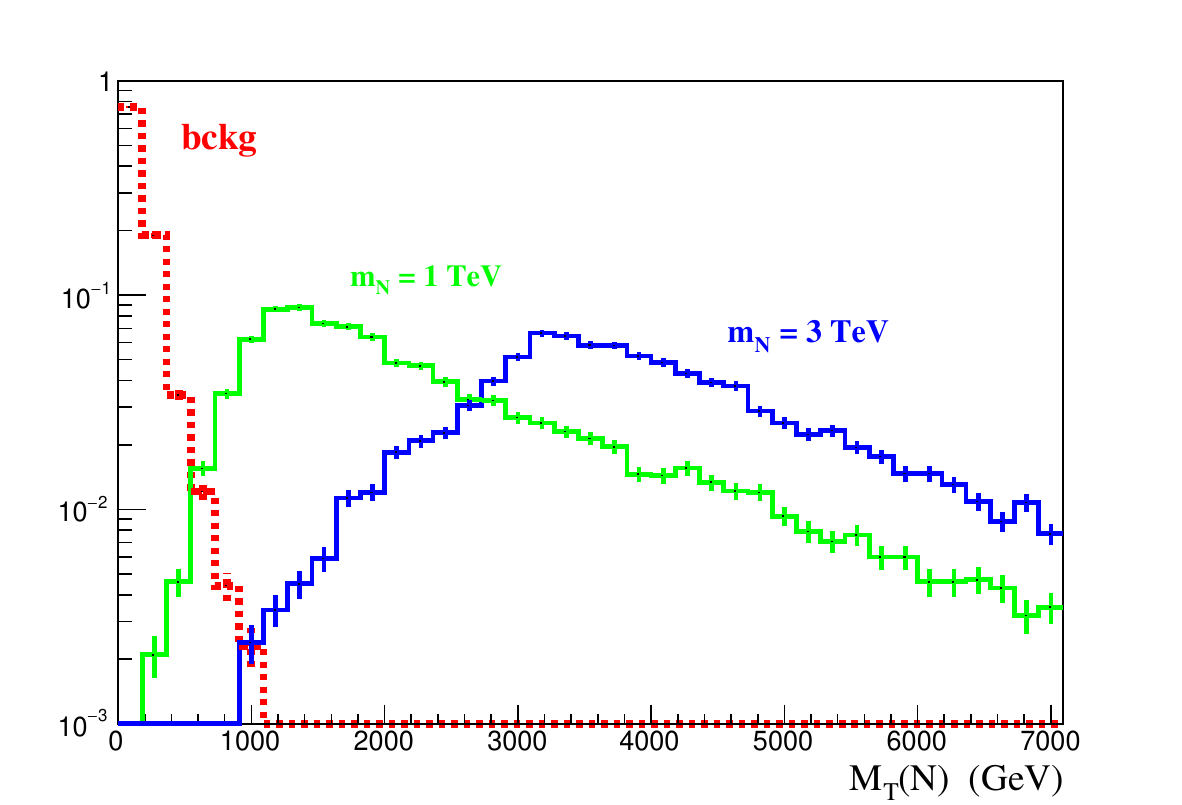} \includegraphics[scale=0.35]{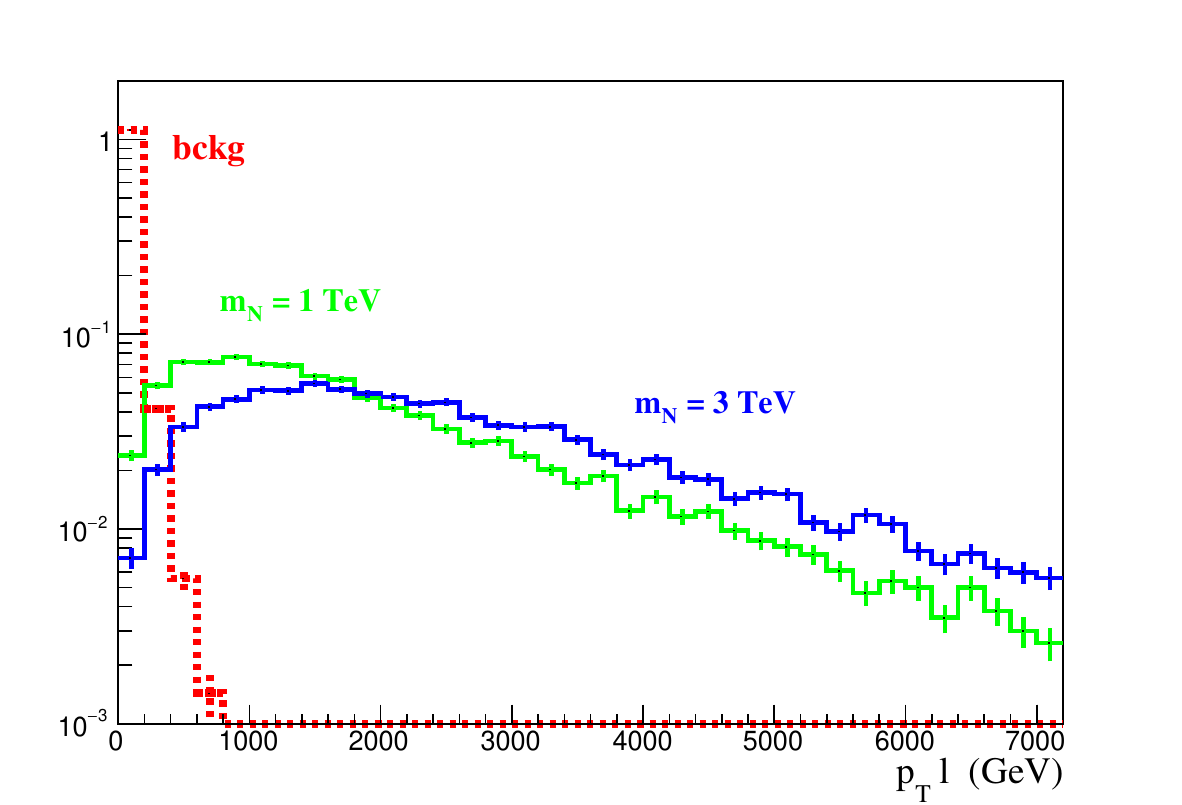}
 \caption{\small \em Distributions normalized to unit area for the $pp\to \ell^\mp \gamma jj+E^{miss}$ VBF dipole-portal signal in scenario (D) $d_W=0$ with different $N$ masses, and for the SM background. We plot the transverse mass of the sterile neutrino (left plot) and the transverse momentum of the lepton (right plot).
 }\label{fig:dist-GammaLnu}
 \end{figure}

\paragraph{(iii) $\bf{\ell^\pm \gamma jj}+E^{miss}$\\ \noindent} 
For the theoretical scenarios (C) and (D), we focus on the VBF $N$ production associated with a lepton followed by the decays $N\to \gamma\nu$. The final state we analyze consists of one high energetic lepton, one high energetic photon, at least two jets, and large missing energy. The acceptance requirements are: $p_T>20$ GeV for the transverse momenta of all the final state particles, $|\eta|<2.5$ for the lepton and photon, and $|\eta|<5$ for the jets. Photon, lepton and jets must be separated by $\Delta R>0.4$. The leading SM background is given by $W_\ell\gamma+$jets events, where the jets (at least two) are produced by QCD or EW interactions. The latter include VBF $W\gamma$ production. After acceptance requirements, the background cross section is of 67 pb. As for the other VBF channels, we select our signal by first applying the VBF cuts in Eq.~\eqref{eq:VBF-cut} and then we refine the analysis by exploiting the kinematic characteristics of the signal (cf. Fig.~\ref{fig:dist-GammaLnu}). In particular, we apply the cuts:
\[
p_T\ell>200\,\text{GeV}, \; p_T\gamma>200\,\text{GeV}, \;  E^{miss}_T>200\,\text{GeV}, \;M_T(N)>90\%\cdot m_N\, ,
\]
where $M_T(N)$ denotes the transverse mass of the sterile neutrino, which is defined as  
\[
M_T(N)\equiv\sqrt{2 \, E^{miss}_T\, p_T\gamma\, \left(1-\Delta\phi(E^{miss},\gamma)/\pi \right)} \, .
\]
Signal efficiencies to this selection are higher than 70\%, while the background is reduced to less than 15 fb.

\begin{figure}[h]
\centering
 \includegraphics[scale=0.55]{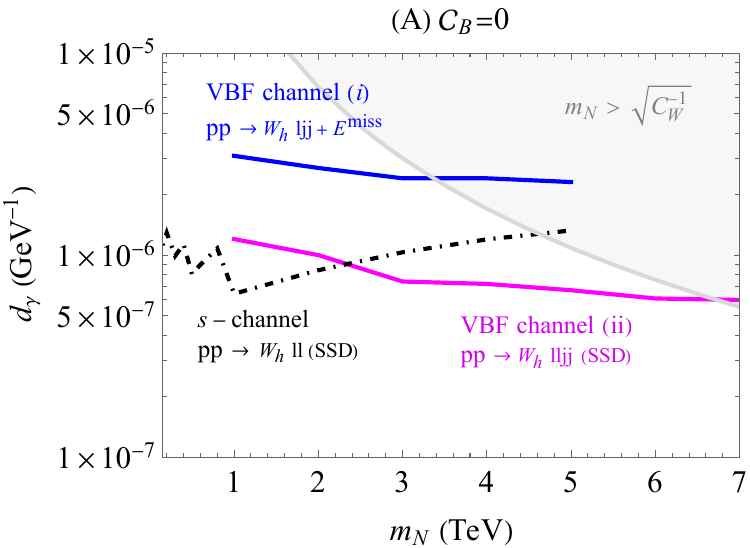} \hspace{0.2cm} \includegraphics[scale=0.55]{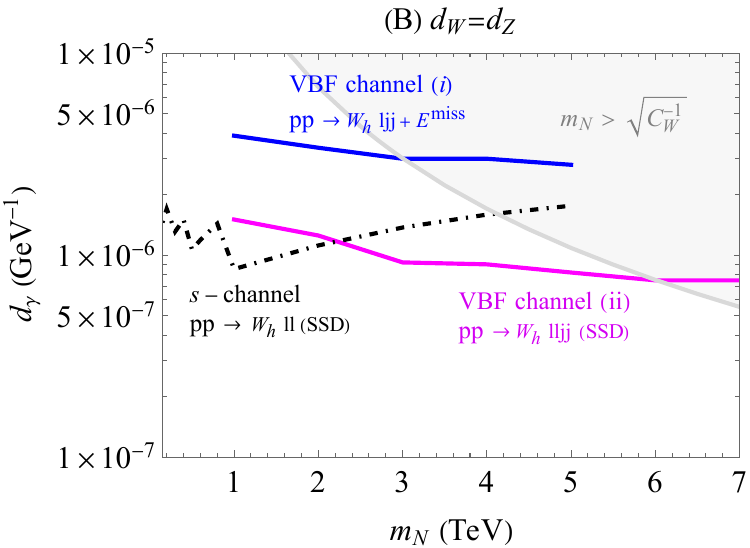} \\ \vspace{0.2cm}
  \includegraphics[scale=0.55]{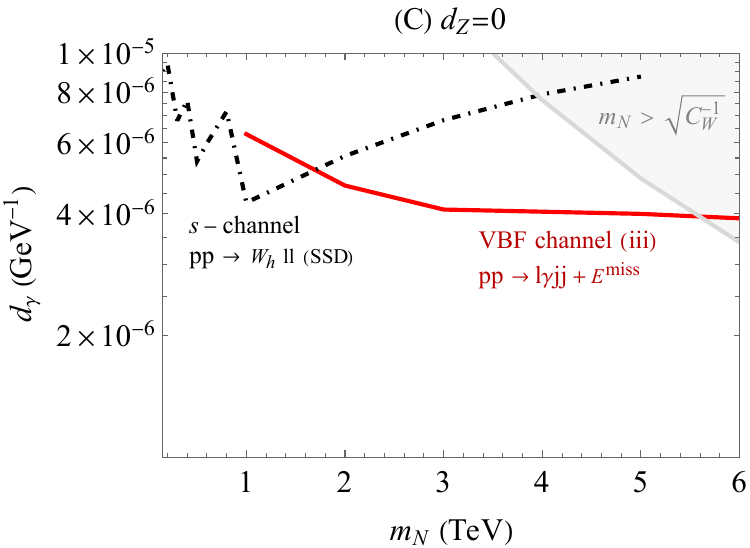} \hspace{0.2cm}\includegraphics[scale=0.55]{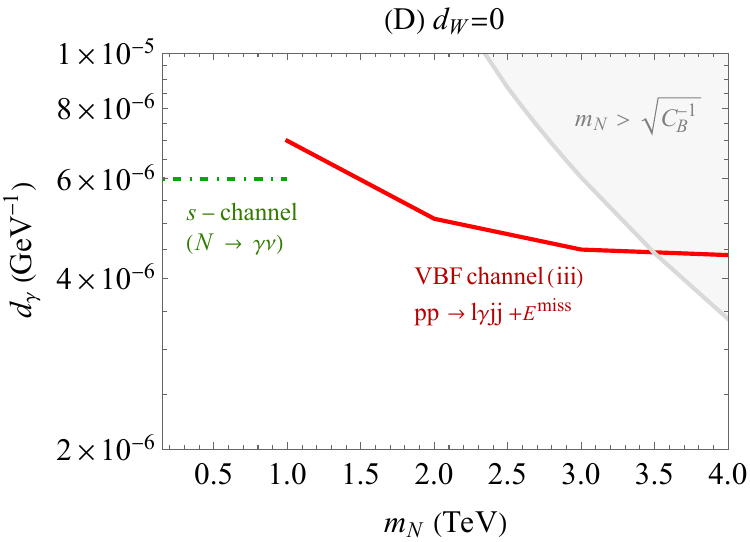}
 \caption{\small \em 2$\sigma$ sensitivities of FCC-hh to the dipole operator $d_\gamma$ for the various theoretical benchmarks considered and in the different analyzed channels.
 }\label{fig:FCC-sensitivities}
 \end{figure}

 At the end of our analyses in the various channels we find the 2$\sigma$ sensitivities of FCC-hh to the dipole operator $d_\gamma$ shown in Fig.~\ref{fig:FCC-sensitivities} for the different theoretical scenarios (A)-(D). The projected sensitivities equally refer to a muonic or electronic flavor\footnote{Possible differences due to different identification efficiencies of muons or electrons are negligible for our purposes.} (we recall that we have assumed a universal flavor scenario). We show the sensitivities corresponding to each of the channels analyzed. In the case of the $s$ channel for the benchmark (D), the reported sensitivity is the one extrapolated from the results of the recent literature \cite{Magill:2018jla,Brdar:2025iua} considering $N\to \gamma\nu$ decays. The shaded regions in the upper right are indicative of a parameter space where the effective-field-theory description cannot be valid (the sterile neutrino mass is indeed above the energy scale of the Wilson coefficients).

 The sensitivity of the $s$-channel $pp \to W_h^\pm\ell^\mp\ell^\mp$ is of the order of 5$\times$10$^{-6}$ GeV$^{-1}$ in scenario (C) and improves to values slightly below 1$\times$10$^{-6}$ GeV$^{-1}$ in benchmarks (A) and (B). For sterile neutrino masses greater than about 2 TeV, we can see how VBF channels offer the best opportunities to probe $d_\gamma$. For scenarios (C) and (D), the study of the channel $pp\to \ell^\pm \gamma jj+E^{miss}$ allows testing dipole operators of the order of 4$\times$10$^{-6}$ GeV$^{-1}$ up to masses $m_N$ of about 5 and 3.5 TeV, respectively. The VBF channel $pp\to W_h^\pm \ell^\mp \ell^\mp jj$ presents the highest sensitivities, that reach $d_\gamma$ of the order of 6-7$\times$10$^{-7}$ GeV$^{-1}$ up to masses $m_N$ of about 6 TeV for scenarios (A) and (B).
Finally, as already underlined, it should be noted that the most efficient channels also offer the possibility of gaining information on the flavor of the dipole operator, which would be obtained by tagging the leptonic flavor.

\section{Muon collider}~\label{sec:mucol}

We reconsider the sensitivity of the muon collider to the dipole portal, highlighting the importance of VBF heavy neutrino production channels that have been overlooked in recent studies~\cite{Barducci:2024kig, Brdar:2025iua}. 

The singular advantage of VBF channels is that they allow significant N production at the muon collider even for flavors other than muonic.
In addition, in the cases with $\mathcal{C}_W\neq0$, we will show that it is necessary, in order to preserve EW gauge invariance and for a correct evaluation of the N production processes, to include all effective interaction terms, in particular those with two gauge bosons from the expansion of $\mathcal{O}_W$. 

We will consider a muon collider with a collision energy $\sqrt{s}=$10 TeV and an integrated luminosity of 10 ab$^{-1}$, which is expected to be collected at the end of a $\sim$5 year run \cite{Accettura:2023ked}.

\subsection{Sterile neutrino production at a muon collider}
\label{sec:production-mucol}

\begin{figure}[]
\centering
 \includegraphics[scale=0.55]{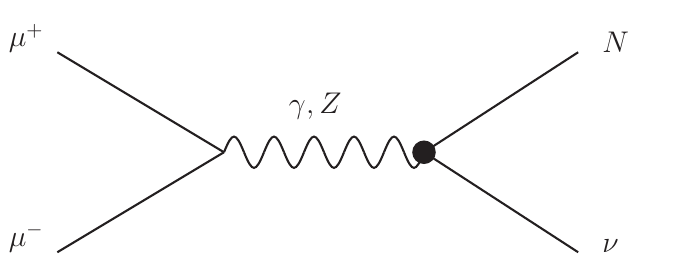} \includegraphics[scale=0.55]{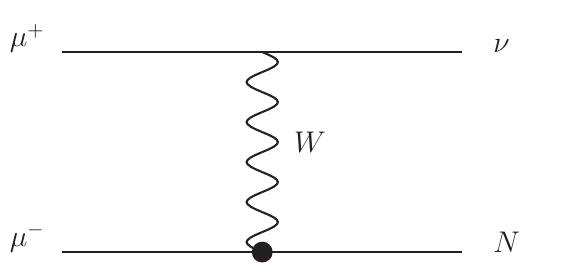}
 \caption{\em \small Feynman diagrams for the $N$ production at a muon collider in \textbf{2-2 Processes}. Left diagram: $s$-channel production induced by the $\bar{\nu}\sigma^{\mu\nu}N F_{\mu\nu}$ or the $\bar{\nu}\sigma^{\mu\nu}NZ_{\mu\nu}$ dipole operator, whose amplitudes are proportional to the coupling $d_\gamma$ or $d_Z$ respectively. Right diagram: $t$-channel production induced by the $\bar{\ell}_L\sigma^{\mu\nu}N W_{\mu\nu}$ dipole operator and with amplitude proportional to $d_W$.}\label{fig:2-2-mucol-dia}
 \end{figure}

\begin{figure}[]
\centering
 \includegraphics[scale=0.48]{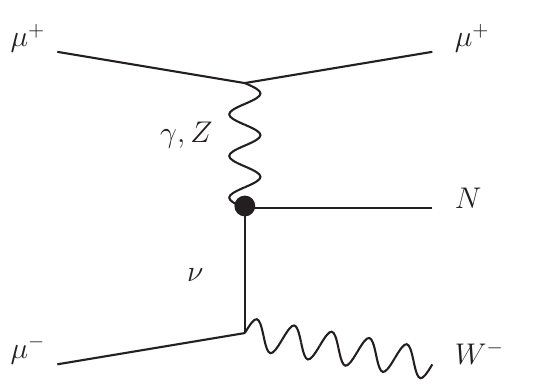} \hspace{0.2cm}
 \includegraphics[scale=0.48]{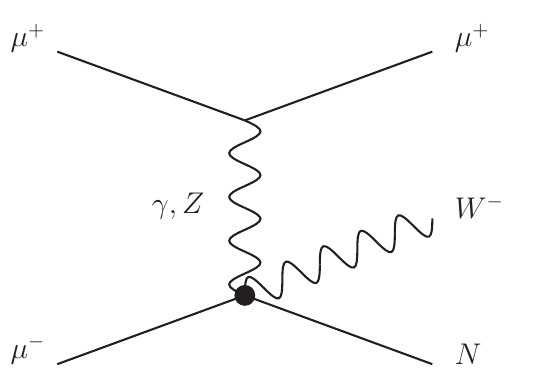} \\ \vspace{0.2cm}
  \includegraphics[scale=0.48]{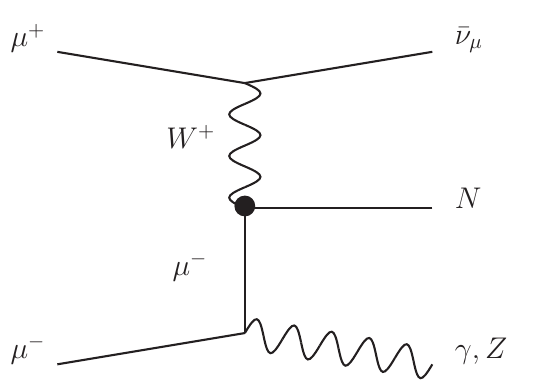} \hspace{0.2cm}
 \includegraphics[scale=0.48]{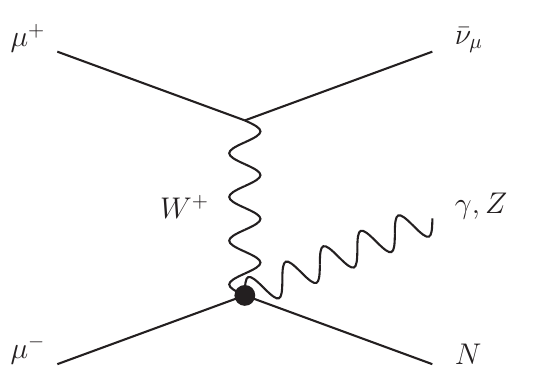}
 \caption{\em \small Leading Feynman diagrams for the $N$ production at a muon collider in \textbf{2-3 Processes}. Upper diagrams: $\mu^+ \mu^- \to N W \mu$. Lower diagrams: $\mu^+ \mu^- \to N \gamma\nu, N Z\nu$. The diagrams on the left are generated by induced dipole effective interactions with a single boson (with couplings $d_\gamma$, $d_Z$ or $d_W$, depending on the specific boson). The diagrams on the right are induced by effective interactions with two gauge bosons, emerging from the expansion of the dipole operator $\mathcal{O}_W$, with amplitudes proportional to $d_W$.}\label{fig:2-3-mucol-dia}
 \end{figure}

\begin{figure}[]
\centering
 \includegraphics[scale=0.48]{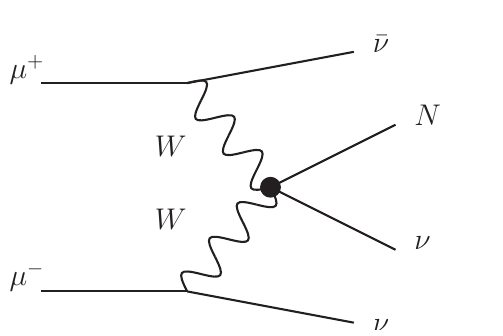} 
 \includegraphics[scale=0.48]{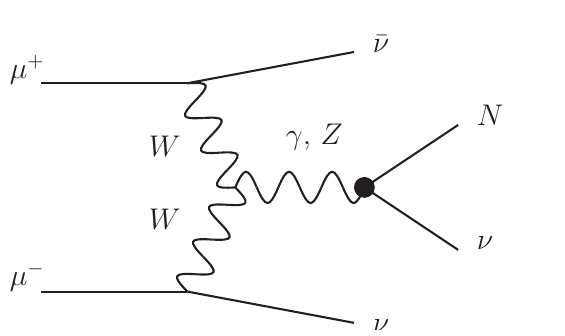}
 \includegraphics[scale=0.45]{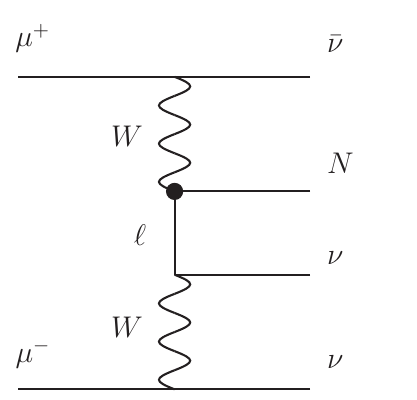} \\ \vspace{0.5cm}
\includegraphics[scale=0.48]{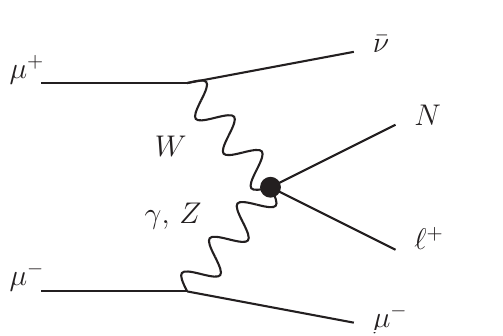} 
 \includegraphics[scale=0.48]{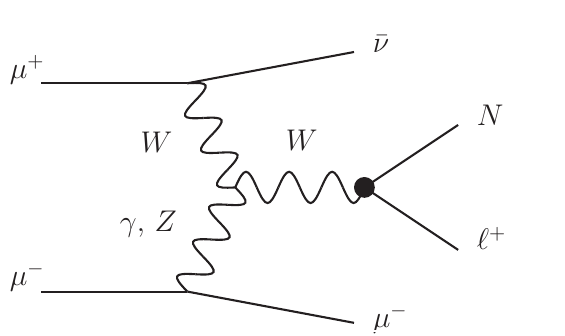}
 \includegraphics[scale=0.45]{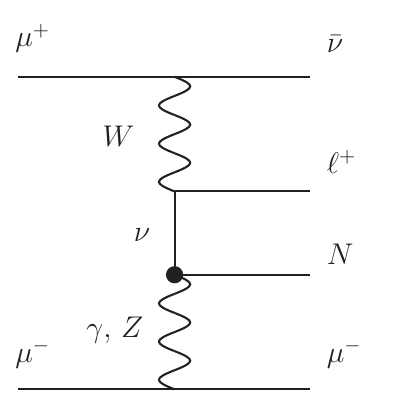}
 \includegraphics[scale=0.45]{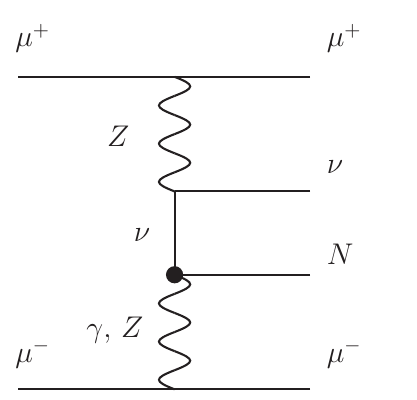}
 \caption{\em \small Representative Feynman diagrams for the $N$ production at a muon collider in \textbf{VBF 2-4 Processes}. Upper diagrams: ``neutral" channels $\mu^+ \mu^- \to N \nu \nu_\mu \bar{\nu_\mu}$. Lower diagrams: ``charged" channels $\mu^+ \mu^- \to N \ell \mu \nu$. Note that the latter include $\mu^+ \mu^- \to N\nu \mu^+ \mu^- $ events (last diagram). }\label{fig:vbf-mucol-dia}
 \end{figure}

The $N$ production in a muon collider can proceed through several processes. The most relevant ones can be classified into three categories: the 2-2 process $\mu^+ \mu^- \to N\nu$, 2-3 processes induced by the exchange in the $t$-channel of a gauge boson $W,Z$ or $\gamma$, and the VBF production, which consists of 2-4 processes induced precisely by the exchange of two gauge bosons in the $t$-channel.
In more detail, the 2-2 $\mu^+ \mu^- \to N\nu$ production process, shown in Fig.~\ref{fig:2-2-mucol-dia}, can receive a contribution from the $s$-channel exchange of $\gamma$ or $Z$ bosons, induced by the dipole operators $\bar{\nu}\sigma^{\mu\nu}N Z_{\mu\nu}$ or $\bar{\nu}\sigma^{\mu\nu}NF_{\mu\nu}$, and a contribution from the $t$-channel exchange of a $W$, induced by $\bar{\ell}_L\sigma^{\mu\nu}N W_{\mu\nu}$. Note that while the $s$-channel production is flavor independent, the $t$-channel production is only possible for the muon flavor. Generally, the contribution to the $t$-channel cross section is dominant. We find that contribution from $t$-channel $W$ exchange is about two orders of magnitude larger than $s$-channel ones in scenarios (A) and (B), and one order of magnitude in scenario (C). Obviously, the $t$-channel contribution is absent for scenario (D) $d_W=0$. 
The production in 2-3 processes is shown in Fig.~\ref{fig:2-3-mucol-dia}. Production in both the neutral channels $\mu^+ \mu^- \to N\gamma\nu, NZ\nu$ and the charged channel $\mu^+ \mu^- \to NW\mu$ is possible. Again, $N$ production is only possible for the muonic flavor. Note that, except for the case $\mathcal{C}_W=0$ ($d_W=0$), there are two main contributions to 2-3 processes, the one induced by single-boson effective interactions (left diagrams in Fig.~\ref{fig:2-3-mucol-dia}) and the one induced by effective interactions with two gauge bosons, generated by the expansion of the operator $\mathcal{O}_W$ (right diagrams in Fig.~\ref{fig:2-3-mucol-dia}). Both of these contributions must be included for a correct gauge-invariant treatment. 
In fact, if two-boson effective interaction diagrams are neglected, the evaluation of 2-3 processes (with the exception of the benchmark (D) $d_W=0$) shows, as an effect of the violation of the EW gauge invariance, an apparent but not consistent increase in the respective cross sections.
Finally, $N$ can be produced, similar to what was previously considered in hadron colliders, via VBF. Such production is possible for any flavor, not just the muon. The 2-4 VBF processes, of which we show representative diagrams in Fig.~\ref{fig:vbf-mucol-dia}, consist of both neutral $\mu^+ \mu^- \to N\nu\nu_\mu\bar{\nu_\mu}$ and charged $\mu^+ \mu^- \to N\ell\mu\nu$ channels. Note that the latter also include $\mu^+ \mu^- \to N\nu\mu^+\mu^-$ events, generated from the scattering of two neutral bosons (last diagram in Fig.~\ref{fig:vbf-mucol-dia}).\footnote{This is an important difference with the case of VBF production at hadron colliders, where $N$ production accompanied by a lepton can only occur by $WV$ fusion ($V\equiv Z,\gamma$), while $VZ$ scattering contributes to the lepton-free final state. For this reason, at the muon collider the ``charged" VBF production is generally more relevant than the ``neutral" one.} 
VBF production at a muon collider depends on a delicate balance between numerous contributions involving single- and double-boson effective vertices and different dipole operators. Also in this case it is crucial to include all possible contributions in order to correctly evaluate, in a gauge-invariant way, the processes.

\begin{figure}[h]
\centering
 \includegraphics[scale=0.42]{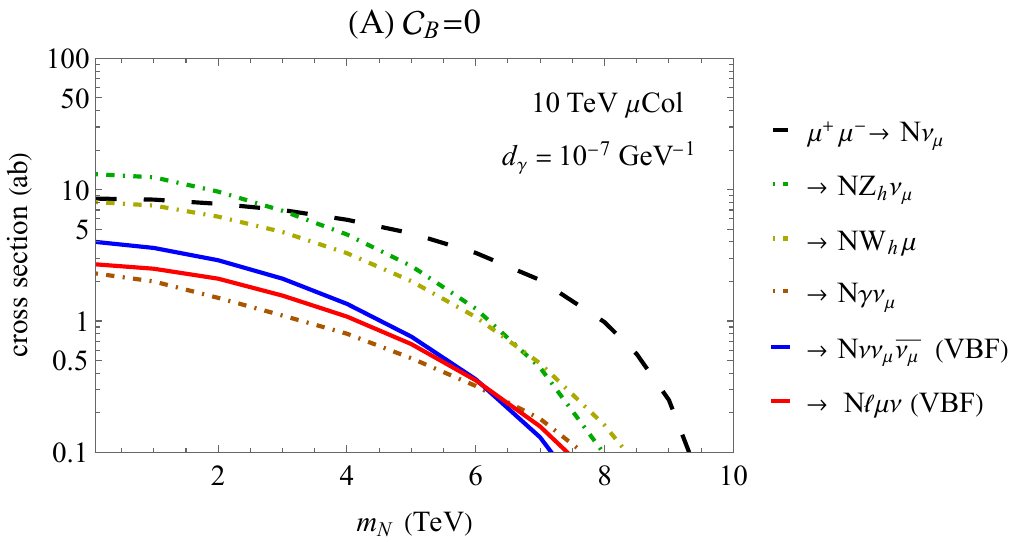} \includegraphics[scale=0.42]{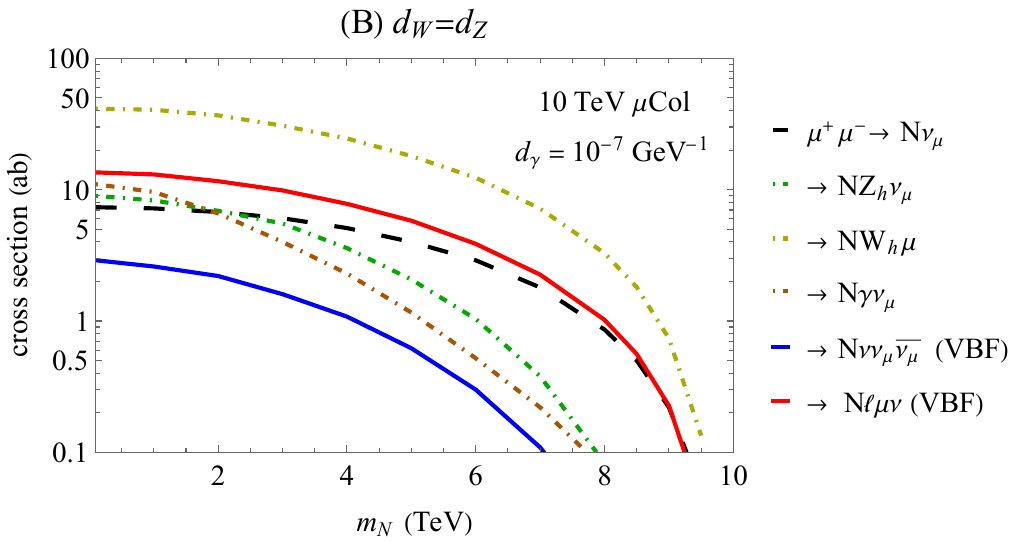}\\ \vspace{0.2cm}
 \includegraphics[scale=0.42]{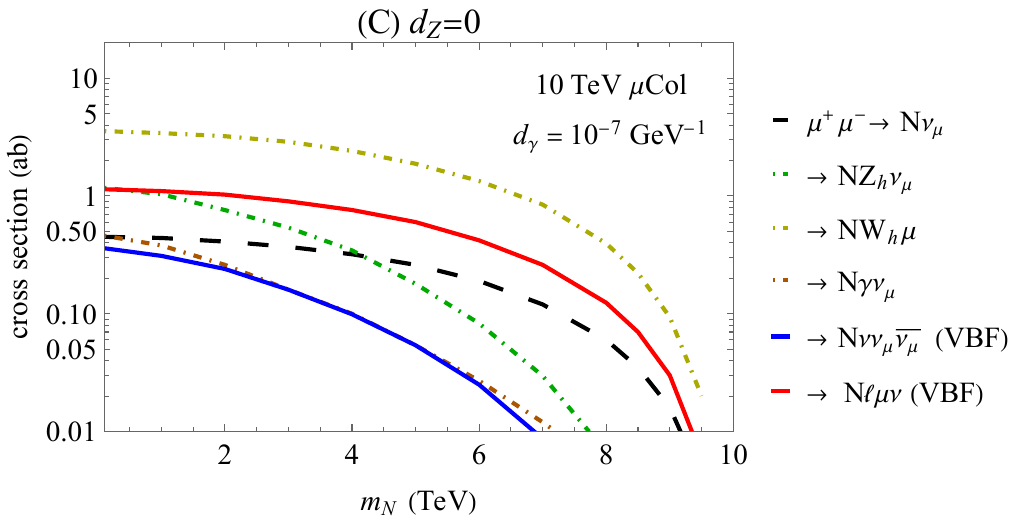} \includegraphics[scale=0.42]{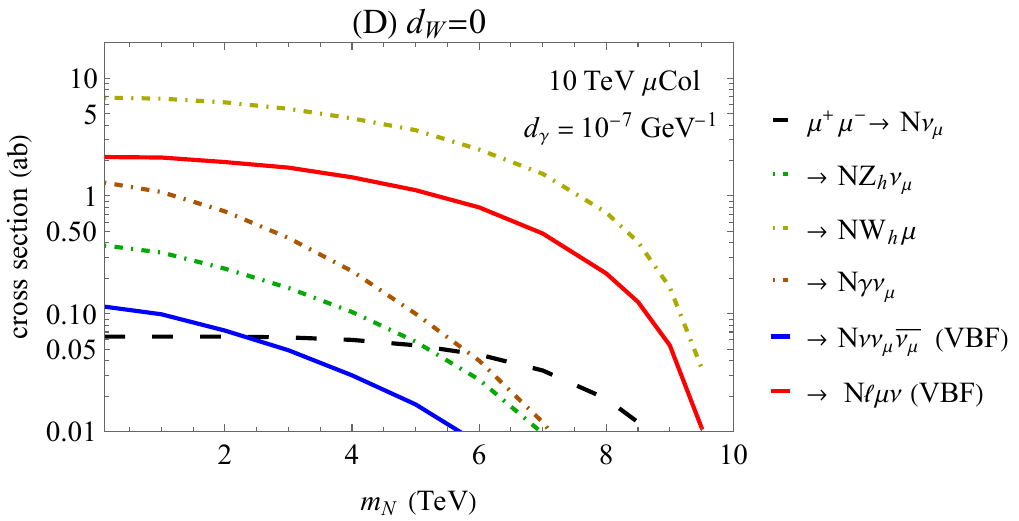}
 \caption{\small \em Cross section for the $N$ production at a muon collider with $\sqrt{s}=10$ TeV in the different theoretical benchmarks and channels: 2-2 process $\mu^+ \mu^- \to N\nu$ (black-dashed curve), 2-3 processes $\mu^+ \mu^- \to N\mu W_h,\, N\nu Z_h,\, N\nu\gamma $ (dot-dashed curves) and VBF 2-4 channels (continuous curves). Cross sections are shown for a reference value $d_\gamma=10^{-7}$ GeV$^{-1}$ as a function of the $N$ mass. 
 }\label{fig:xsec-MuCol}
 \end{figure}

We show in Fig.~\ref{fig:xsec-MuCol}  cross sections for the $N$ production in the different processes and theoretical benchmarks. We consider hadronically decayed bosons in 2-3 processes $\mu^+ \mu^- \to N\mu W_h,\, N\nu Z_h$. 
We find that flavor-independent VBF production is relevant in all theoretical scenarios considered and generally exhibits higher cross sections, as expected, in scenarios with higher $\mathcal{C}_W$ values.
The lepton-associated VBF production  $\mu^+ \mu^- \to N \ell \mu \nu$ is particularly notable, constituting the production channel with the highest cross section, after $\mu^+\mu^-\to NW_h\mu $, in scenarios (B), (C) and (D).

 The 2-2 process is, with the exception of scenario (D), one of those with the highest cross section. The 2$\sigma$ sensitivity to the $d^{(\mu)}_\gamma$ operator in this channel has been evaluated in \cite{Brdar:2025iua, Barducci:2024kig}, obtaining values in the range 1-2 $\times 10^{-6}$ GeV$^{-1}$ in scenario (C) $d_Z=0$.
 In scenarios (B), (C), and (D), the 2-3 process $\mu^+ \mu^- \to N\mu W_h,$ has the highest cross section. In scenario (D), the 2$\sigma$ sensitivity to the $d^{(\mu)}_\gamma$ operator in this channel evaluated in \cite{Barducci:2024kig} is approximately $1\times 10^{-7}$ GeV$^{-1}$. For the other theoretical benchmarks, however, the channel sensitivity should be re-evaluated through a gauge-invariant treatment, which we defer to future studies.
 In the present work, we instead want to focus on flavor-independent VBF channels.
Depending on the decays of $N$, we can distinguish the following final states:

\begin{multicols}{2}
    \begin{enumerate}[label=(\roman*)]
        \item $(N\to W^\pm \ell^\mp)\nu \nu_\mu \bar{\nu}_\mu \to  W^\pm \ell^\mp +E^{miss} $
           \item $(N\to W \ell)\ell \mu \nu \to W \ell \ell \mu +E^{miss}$ 
          \item $(N\to \gamma\nu)\ell^\pm \mu^\mp \nu \to \gamma \ell^\pm\mu^\mp+E^{miss} $
           \item $(N\to \gamma\nu)\nu \nu_\mu \bar{\nu}_\mu  \to \gamma +E^{miss} $
           \item$(N\to Z\nu)\nu \nu_\mu \bar{\nu}_\mu  \to Z +E^{miss} $
          \item$(N\to Z\nu)\ell^\pm \mu^\mp \nu \to Z\ell^\pm\mu^\mp+E^{miss} $
    \end{enumerate}
    \end{multicols}

Once again, a braided analysis in all available channels would be ideal. For the sake of concreteness, in the present study we will focus only on the most promising channels in each of the theoretical scenarios considered. Taking into account the cross section trends and branching ratios for the various benchmarks, we will analyze the channels (i) and (ii ) for scenarios (A) and (B). Note that channel (ii) offers an opportunity (similarly to what we have already seen for hadronic colliders) to identify the flavor of the dipole operator involved in both the production~\footnote{Except where $N$ is produced in $\mu^+ \mu^- \to N\nu \mu^+ \mu^-$ channel from $ZV$ scattering.} and the decay of the sterile neutrino. Due to the higher BRs in $\gamma\nu$ for the benchmarks (C) and (D), we will consider channel (iii). 
For channels (i) and (ii) involving a $W$, we will consider that it decays hadronically. As already highlighted in several previous studies \cite{Vignaroli:2023rxr, Barducci:2024kig,Brdar:2025iua}, this $W$ is highly boosted and can therefore be easily reconstructed as a single jet. For simplicity and similar to what was done in \cite{ Barducci:2024kig, Brdar:2025iua}, we will not include in our simulations the decay with subsequent hadronization and showering of the $W$, but we will take it into account by multiplying the signal and background yields by a factor BR$(W\to jj)\approx 0.67$.

 \subsection{Search strategies and sensitivity projections in VBF channels}\label{sec:mucol-sensitivities}

We outline signal selection strategies for the most promising channels, deriving sensitivity estimates of the muon collider to the sterile neutrino-dipole portal in the different benchmarks.

 \paragraph{(i) $\bf{W_h^\pm \ell^\mp +E^{miss}}$ \\}
 We focus on a final state consisting of exactly one lepton plus a hadronically decayed $W$ and missing energy. This channel is particularly promising for the test of scenarios (A) $\mathcal{C}_B=0$ and (B) $d_W=d_Z$.
 After imposing the acceptance requirements on the lepton and $W_h$ transverse momenta and pseudo-rapidities: $p_T>20$ GeV, $|\eta| <2.5$, we find a total SM background at the level of 270 fb, dominantly given by $W_h \mu \nu_\mu$ events and subdominantly by $W_h W_\ell$ production. The latter, which includes both electrons and muons from the leptonic decay of $W_\ell$ has a cross section of about 4 fb. 
 The signal is characterized by the hard emission of the lepton and the $W_h$ and by large missing energy. We also observe that the signal lepton is emitted more centrally than in the background events. The sterile neutrino can be easily reconstructed by combining the lepton and the $W_h$. We thus apply the following signal selection cuts:
\[
|\eta(\ell)|<2 \,, \; p_T \ell>200 \, \text{GeV}, \; E^{miss}_T>200 \, \text{GeV}, \; p_T W_h>200 \, \text{GeV}, \; M_{W_h\ell} \in [0.8,1.2]\cdot m_N \, ,
\]
which reduce the background cross section below $\sim$5 fb. 
We can refine the analysis for the case $N\to e W_h$. If we restrict ourselves to the electronic flavor, the background is in fact significantly lower (by approximately two orders of magnitude), receiving a contribution mainly from the subdominant background component $W_h (W \to e\nu_e)$. Applying the same selection cuts as in the generic flavor case, we therefore find higher sensitivities.

\begin{figure}[h!]
\centering
 \includegraphics[scale=0.35]{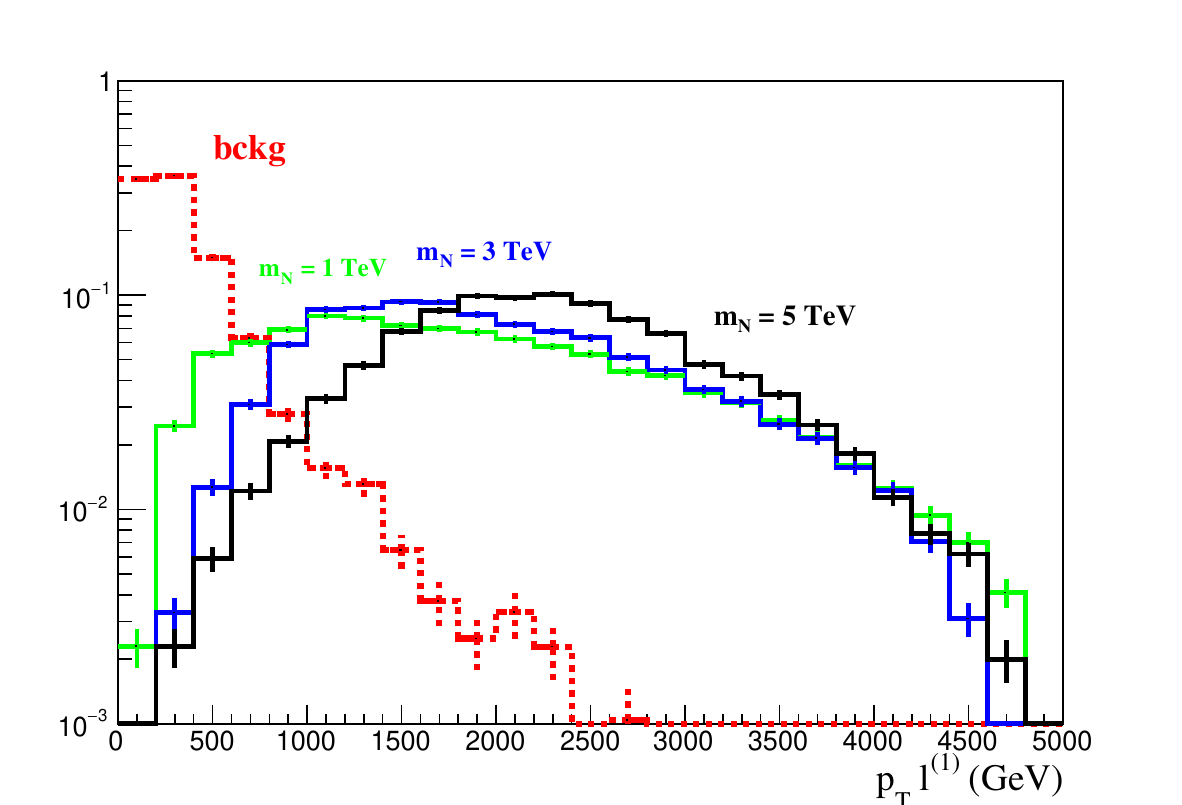} \includegraphics[scale=0.35]{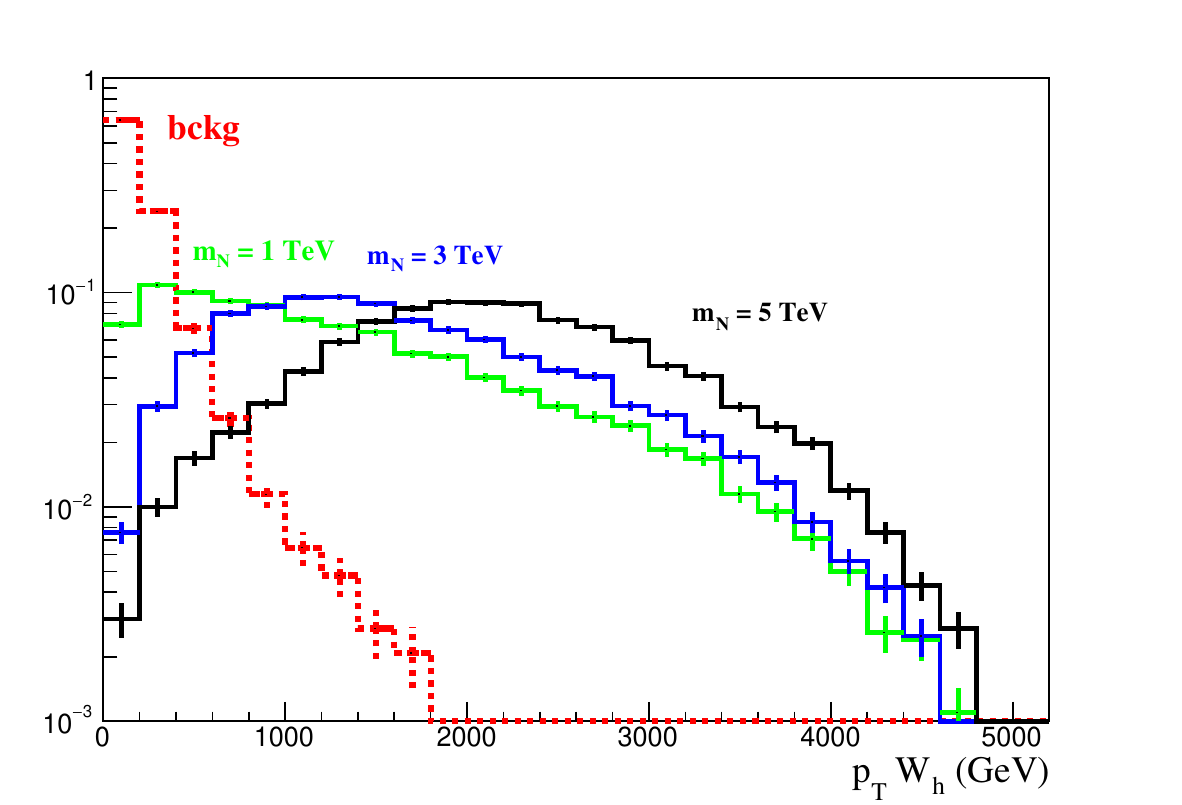}\\
 \includegraphics[scale=0.35]{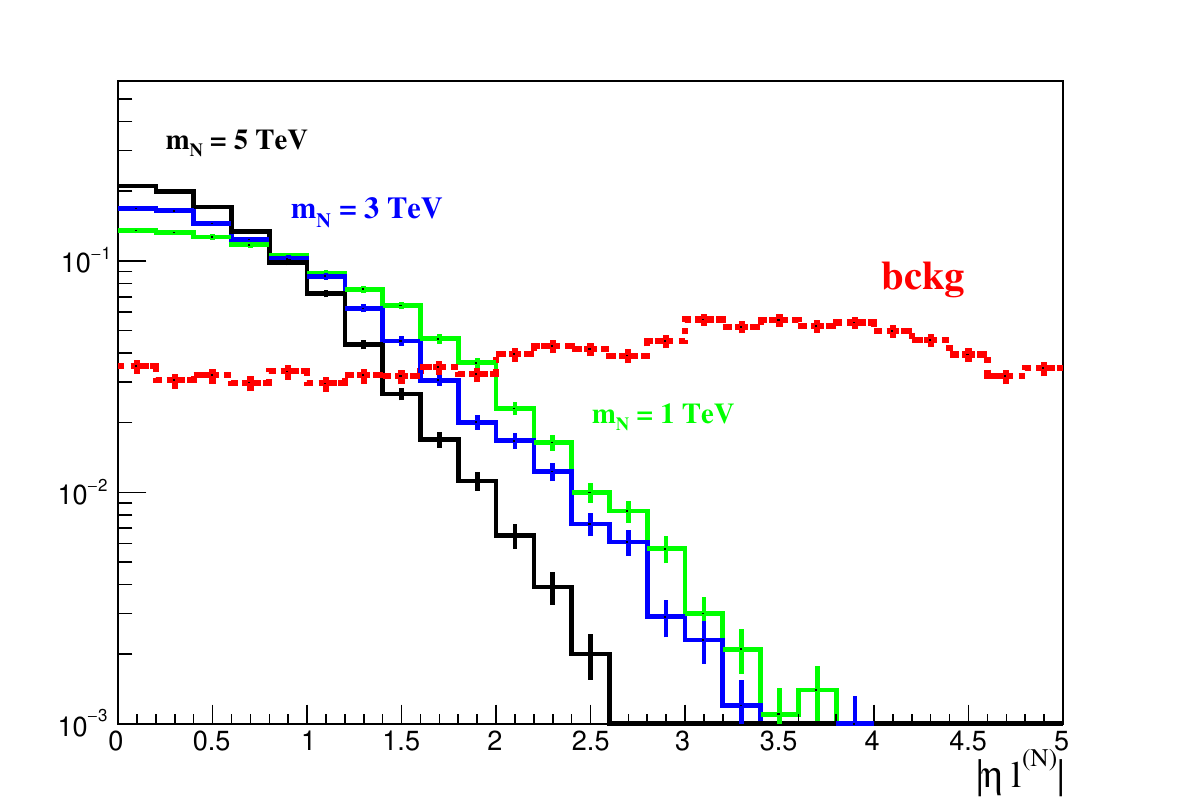} \includegraphics[scale=0.35]{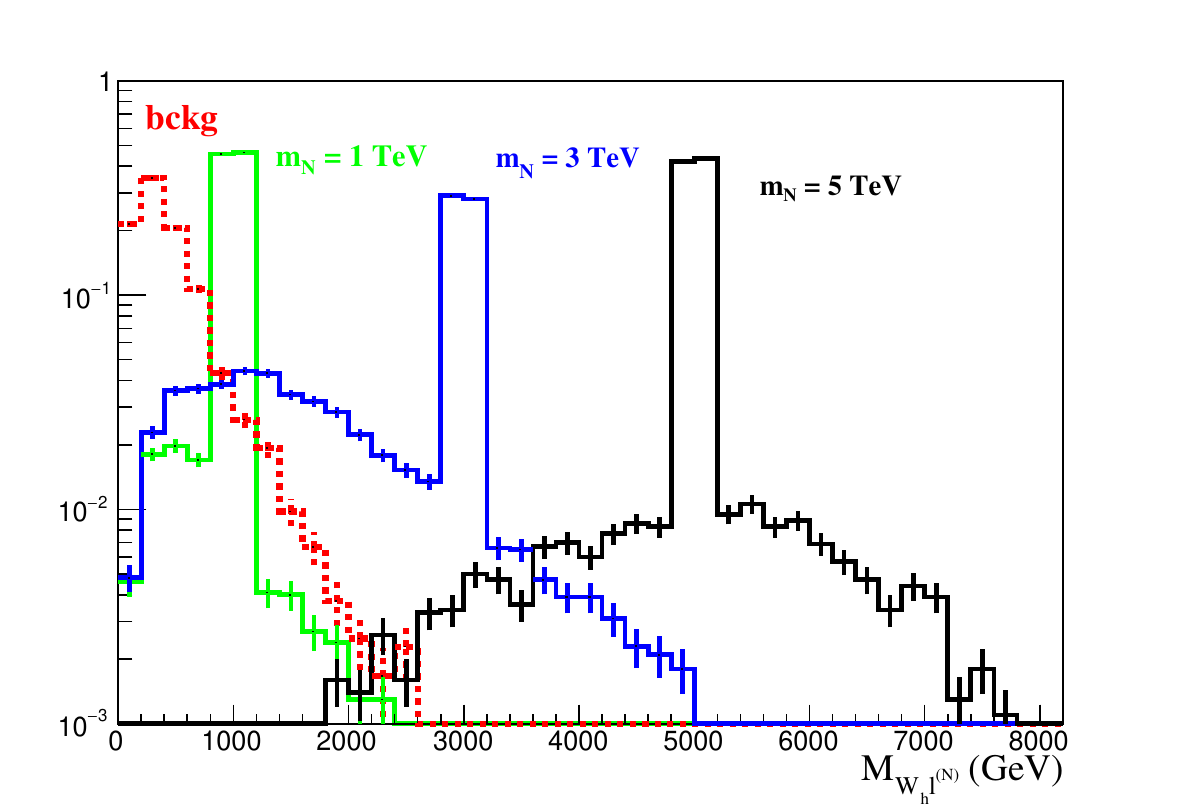}
 \caption{\small \em Distributions normalized to unit area for the $\mu^+ \mu^-\to W_h \ell \ell \mu +E^{miss}$ VBF dipole-portal signal (ii) in the benchmark (B) $d_W=d_Z$ (similar distributions are obtained in scenario (A)) with different $N$ masses, and for the SM background. 
 Upper plots: leading-$p_T$ lepton (left) and $W_h$ (right) transverse momentum distributions. Lower plots: pseudo-rapidity distribution of the identified lepton from $N$ decay, $l^{(N)}$, (left) and invariant mass of the system consisting of $l^{(N)}$ and $W_h$ (right plot).
 }\label{fig:dist-mucol-ii}
 \end{figure}

 \paragraph{(ii) $\bf{W_h \ell \ell \mu +E^{miss}}$ \\}
 In order to more efficiently probe the theoretical benchmarks (A) $\mathcal{C}_B=0$ and (B) $d_W=d_Z$, we also focus on a final state consisting of a $W_h$, two leptons (emitted in the central region $|\eta|<$2.5 with a $p_T>20$ GeV), a muon (with rapidity up to $|\eta|<$5 and $p_T>20$ GeV) and missing energy ($E^{miss}_T>20$ GeV). The SM background, which includes $W_h W_\ell V_\ell$ events, is at the level of 40 fb after acceptance requirements.

As shown in Fig.~\ref{fig:dist-mucol-ii}, signal (ii) is characterized by the presence of the two leptons associated with the production and decay of $N$ emitted at high $p_T$, together with $W_h$, and generally in the central region of the detector. The sterile neutrino can be quite efficiently reconstructed starting from $W_h$ and identifying the lepton coming from its decay in the following way: for the less massive cases, $m_N <5$ TeV, this lepton, which we denote with $\ell^{(N)}$, is for the majority of signal events the one with the smallest separation $\Delta R$ from $W_h$. For cases $m_N\geq 5$ TeV, $l^{(N)}$ instead coincides for the majority of signal events with the leading-$p_T$ lepton $\ell^{(1)}$.
We therefore apply the following selection cuts:  
\[
|\eta(\ell^{(N)})|<2.5 \,, \; p_T \ell^{(1)}>200 \, \text{GeV}, \; p_T W_h>200 \, \text{GeV}, \; M_{W_h\ell^{(N)}} \in [0.8,1.2]\cdot m_N \, .
\]
After these cuts, the background is reduced to less than $\sim$1 fb.

\begin{figure}[h!]
\centering
 \includegraphics[scale=0.35]{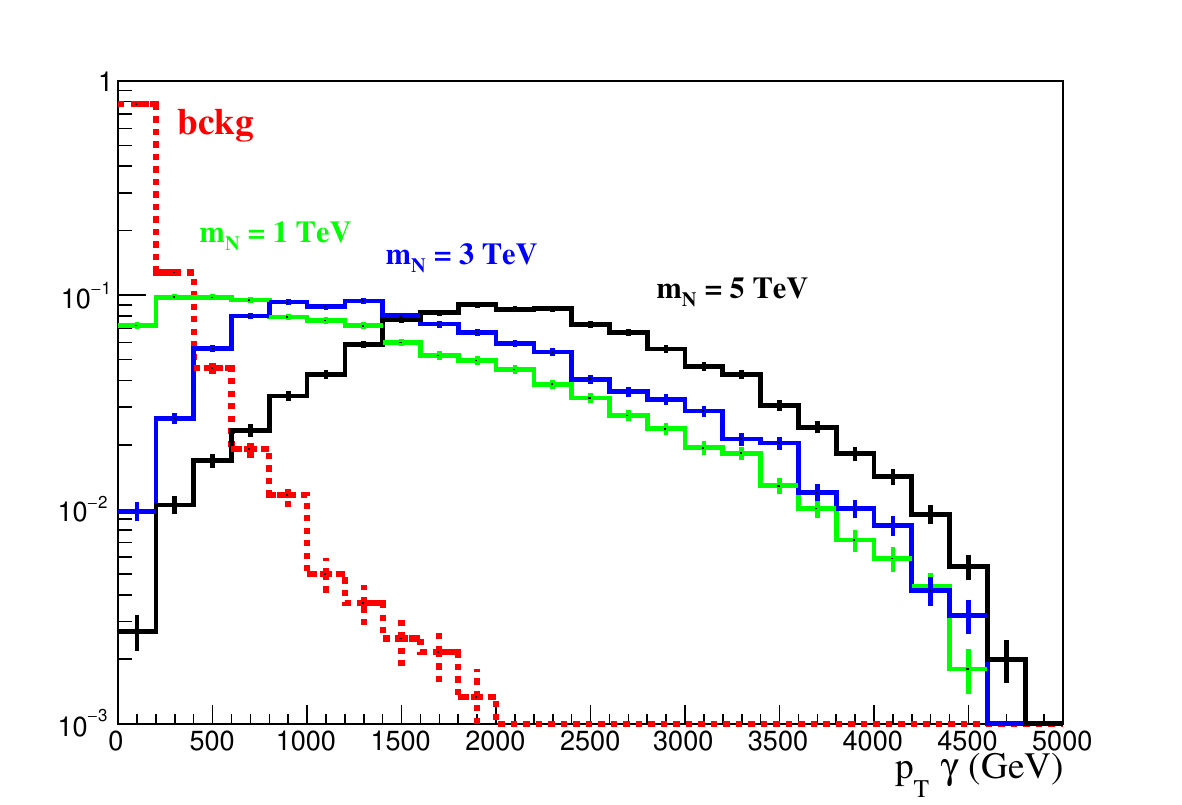} \includegraphics[scale=0.33]{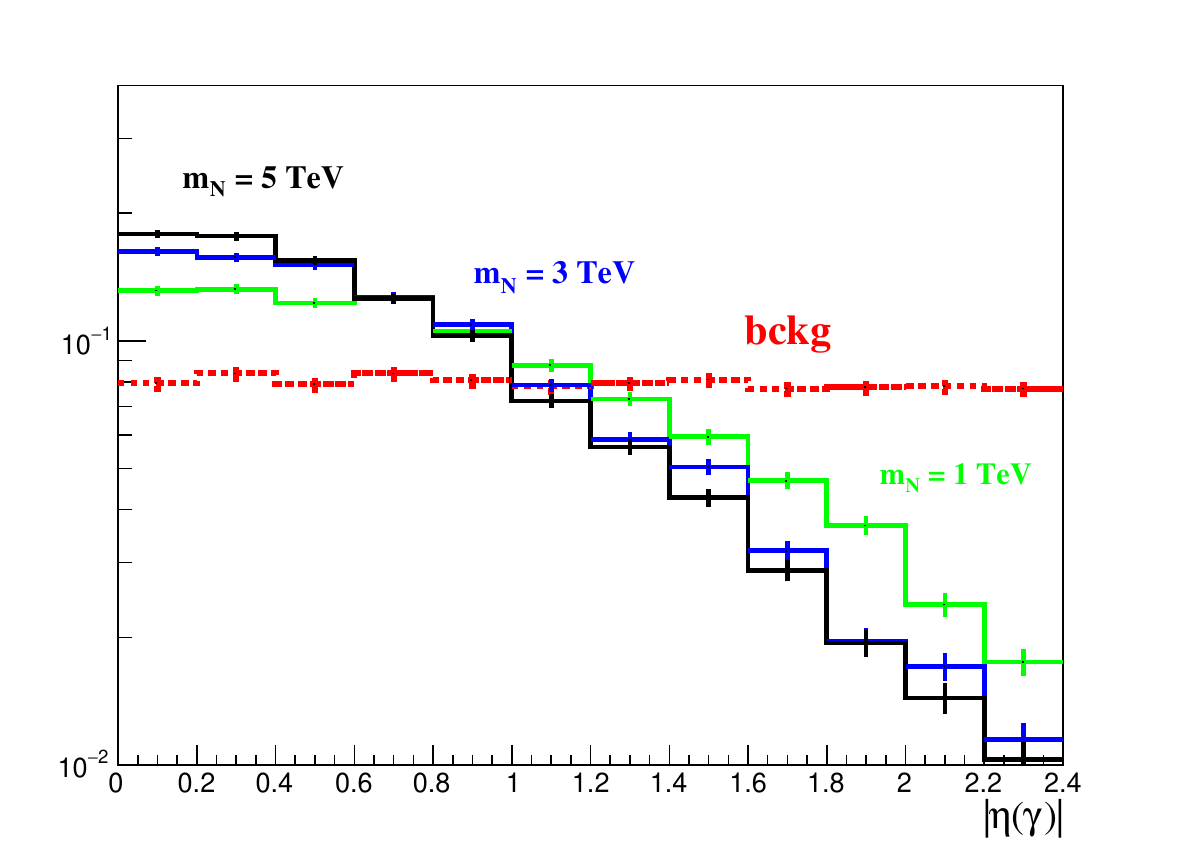}
 \caption{\small \em Distributions normalized to unit area for the $\mu^+ \mu^-\to \gamma\ell \mu +E^{miss}$ VBF dipole-portal signal (iii) in the benchmark (D) $d_W=0$ (similar distributions are obtained in scenario (C)) with different $N$ masses, and for the SM background. 
 We plot the photon transverse momentum (left) and pseudorapidity (right) distributions.  
 }\label{fig:dist-mucol-iii}
 \end{figure}

 \paragraph{(iii) $\bf{\gamma \ell\mu +E^{miss}}$ \\}
We consider channel (iii) to better probe scenarios (C) and (D). In the final state, we require a photon (with $p_T>20$ GeV and $|\eta|<$2.5), two leptons, one of which with muonic flavor (with $p_T>20$ GeV and $|\eta|<$5) and a large missing energy ($E^{miss}_T>20$ GeV). The leptons must be separated from each other and from the photon by a distance $\Delta R>0.4$.
The dominant background consists of the SM production $\gamma \ell \mu \nu \bar{\nu}$, which includes contributions from $\gamma W^+_\ell W^-_\ell$ and $\gamma Z_\ell (Z\to \nu\bar{\nu})$ events. After acceptance requirements, the background cross section is about 40 fb.
The main features of the neutrino dipole-portal signal are the large missing energy and the very energetic photon, emitted much more centrally than the background, as we show in Fig.~\ref{fig:dist-mucol-iii}. We select the signal by applying the following cuts:
\[
|\eta(\gamma)|<1.5 \,, \; p_T\gamma>20\% \cdot m_N, \; \, E^{miss}_T>20\% \cdot m_N.
\]
After these cuts, the background is reduced to less than $\sim$4 fb.

\begin{figure}[h!]
\centering
 \includegraphics[scale=0.55]{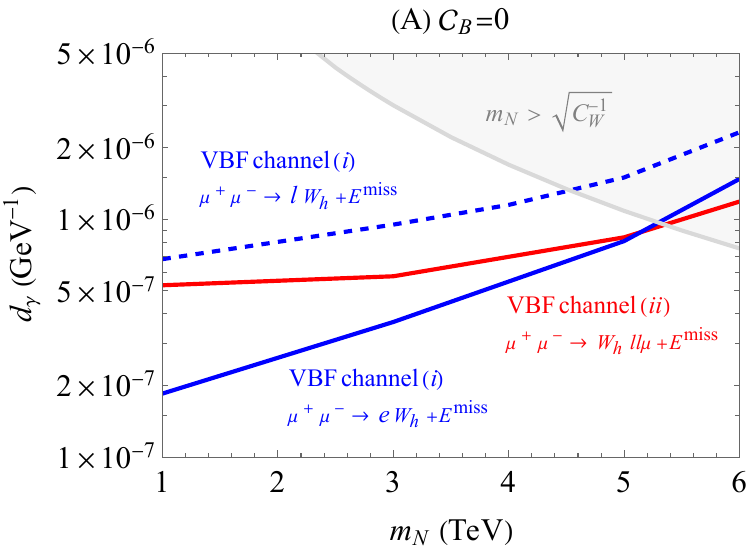} \hspace{0.2cm} \includegraphics[scale=0.55]{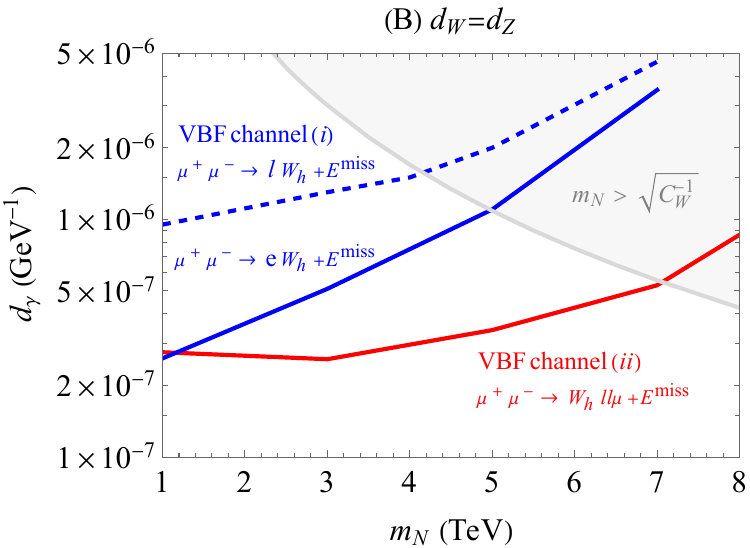} \\ \vspace{0.2cm}
  \includegraphics[scale=0.55]{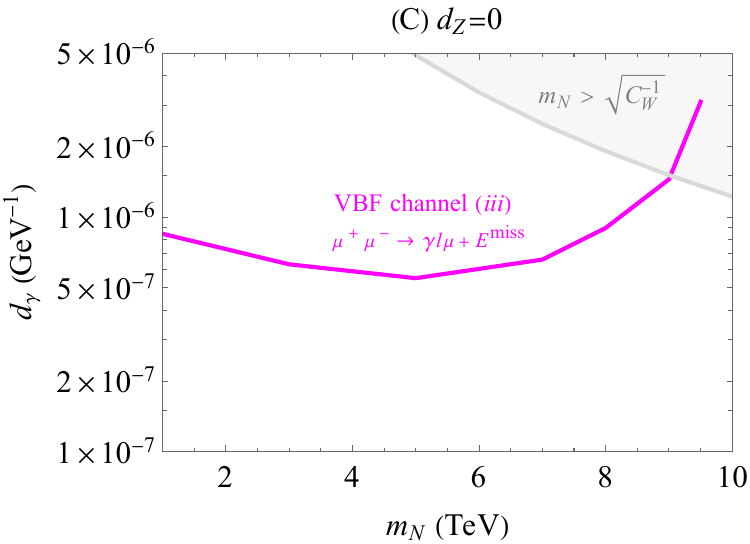} \hspace{0.2cm}\includegraphics[scale=0.55]{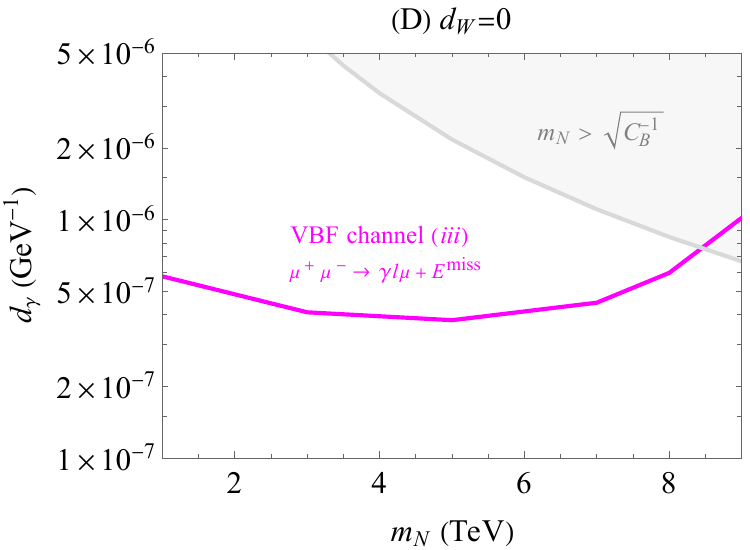}
 \caption{\small \em 2$\sigma$ sensitivities of the 10 TeV muon collider to the dipole operator $d_\gamma$ for the various theoretical benchmarks and in the different analyzed channels.
 }\label{fig:mucol-sensitivities}
 \end{figure}

 Our analyses in the various channels predict the 2$\sigma$ sensitivities of the 10 TeV muon collider to the $d_\gamma$ dipole operator shown in Fig.~\ref{fig:mucol-sensitivities} for the different theoretical benchmarks (A)-(D). Note that the projected sensitivities refer equally to a muonic or electronic flavor (recall that we assumed a flavor-universal scenario). Sensitivity to flavors other than muonic is gained by taking advantage of the unique flavor-independent property of VBF production. Again, the shaded regions indicate a parameter space where the effective field theory description loses its validity.
In scenario (A) $\mathcal{C}_B=0$, in particular from the analysis of channel (i), sensitivities in the range $\sim$2-5$\times$10$^{-7}$ GeV$^{-1}$ are obtained for $N$ up to about 5 TeV. Channel (ii) is especially efficient for probing scenario (B) $d_W=d_Z$, reaching sensitivities of approximately 2.5-5 $\times$10$^{-7}$ GeV$^{-1}$ for $N$ up to $\sim$7 TeV.
Scenarios (C) $d_Z=0$ and (D) $d_W=0$ can be efficiently tested in channel (iii), with sensitivities in the ranges $\sim$5.5-9$\times$10$^{-7}$ GeV$^{-1}$ and $\sim$4-6$\times$10$^{-7}$ GeV$^{-1}$ respectively for masses of $N$ up to $\sim$9 and $\sim$8 TeV.

\section{Results and Outlook}\label{sec:conclusions}

In this work, we have explored novel strategies to probe a neutrino dipole portal at the energy frontier, focusing on the sterile neutrino production and decay signatures across a variety of future collider scenarios.
We considered hadron colliders: the upcoming LHC high-luminosity program and the FCC-hh, and a futuristic muon collider at 10 TeV.

Our dedicated analysis for the High-Luminosity LHC demonstrated that in theoretical scenarios where the neutrino dipole is significantly generated by the $SU(2)_w$ invariant operator $\mathcal{O}_W$, sensitivity to the dipole portal can be significantly improved by considering sterile neutrino decays in $W \ell$, leading to clean same-sign dilepton final states, in the case of Majorana nature of the sterile neutrino. As shown in Figure~\ref{fig:sensitivity-hllhc}, 2$\sigma$ sensitivities to dipole couplings down to $d_\gamma \sim 3 \times 10^{-6}~\text{GeV}^{-1}$ can be achieved, with flavor tagging offering potential insight into the operator structure.

At high energies probed by colliders such as FCC-hh and a muon collider, sterile neutrino production from gauge-boson scattering (VBF) becomes very significant and generally dominant.
For the correct gauge-invariant treatment of these (and other) processes~\footnote{Other important processes which require the inclusion of all the terms arising from $\mathcal{O}_W$ for a correct gauge-invariant treatment (unless in scenario $d_W=0$) are represented by the 2-3 scatterings at a muon collider. In particular, the production channel $\mu^+\mu^- \to N W\mu$ shows a promising cross-section yield.}, it is essential to include effective interactions with two gauge bosons, arising from the non-abelian operator $\mathcal{O}_W$. VBF channels, for which we identified different analysis strategies, can explore regions that are not accessible to other experiments so far. 

At the FCC-hh, our results (Figure~\ref{fig:FCC-sensitivities}) indicate that these channels, particularly the VBF-induced same-sign dilepton and $\ell \gamma jj$ final states, yield improved sensitivities up to $d_\gamma \sim 6 \times 10^{-7}~\text{GeV}^{-1}$ for masses as large as 6~TeV in benchmarks with larger $C_W$ values. Sensitivities are good anyway, $d_\gamma \sim4\times$10$^{-6}$ GeV$^{-1}$ up to about $m_N\sim3.5$ TeV, even in the case where the dipole is instead completely generated by the hypercharge-invariant operator $\mathcal{O}_B$.
This extends the sensitivity frontier significantly beyond previous projections, which focused primarily on $s$-channel production and photon decays.

Most remarkably, our study of the 10~TeV muon collider 
shows that VBF processes provide access to sterile neutrino production even for non-muonic flavors---a critical point overlooked in previous literature. The muon collider achieves the highest sensitivity among all scenarios analyzed, reaching $d_\gamma \sim 2 \times 10^{-7}~\text{GeV}^{-1}$ for the benchmarks with larger $\mathcal{C}_W$ values and remaining at sensitivities of the order $d_\gamma \lesssim 5\times$10$^{-7}$ GeV$^{-1}$ for all the different theoretical benchmarks considered, as depicted in Figure~\ref{fig:mucol-sensitivities}.

In summary, our results highlight the necessity of a fully gauge-invariant treatment of the neutrino dipole portal, including multi-boson operators, to accurately estimate sterile neutrino production at future colliders. Our findings open new directions for the experimental exploration of neutrino portals and demonstrate the significant potential of VBF channels to uncover physics beyond the Standard Model in the neutrino sector.

\section{Acknowledgement}

The author thanks Daniele Barducci for useful discussions and Vedran Brdar for clarifications regarding the study~\cite{Brdar:2025iua}. This work is supported by ICSC – Centro Nazionale di Ricerca in High Performance Computing, Big Data and Quantum Computing, funded by European Union – NextGenerationEU, reference code CN\_00000013.  This study also enjoyed enlightening discussions during the GGI workshop "Exploring the Energy Frontier with Muon Beams".

\bibliographystyle{unsrt}
\bibliography{sample}

\begin{thebibliography}{10}

\bibitem{Minkowski:1977sc}
Peter Minkowski.
\newblock {$\mu \to e\gamma$ at a Rate of One Out of $10^{9}$ Muon Decays?}
\newblock {\em Phys. Lett. B}, 67:421--428, 1977.

\bibitem{Gell-Mann:1979vob}
Murray Gell-Mann, Pierre Ramond, and Richard Slansky.
\newblock {Complex Spinors and Unified Theories}.
\newblock {\em Conf. Proc. C}, 790927:315--321, 1979.

\bibitem{Yanagida:1979as}
Tsutomu Yanagida.
\newblock {Horizontal gauge symmetry and masses of neutrinos}.
\newblock {\em Conf. Proc. C}, 7902131:95--99, 1979.

\bibitem{Mohapatra:1979ia}
Rabindra~N. Mohapatra and Goran Senjanovic.
\newblock {Neutrino Mass and Spontaneous Parity Nonconservation}.
\newblock {\em Phys. Rev. Lett.}, 44:912, 1980.

\bibitem{Fukugita:1986hr}
M.~Fukugita and T.~Yanagida.
\newblock {Baryogenesis Without Grand Unification}.
\newblock {\em Phys. Lett. B}, 174:45--47, 1986.

\bibitem{Akhmedov:1998qx}
Evgeny~K. Akhmedov, V.~A. Rubakov, and A.~Yu. Smirnov.
\newblock {Baryogenesis via neutrino oscillations}.
\newblock {\em Phys. Rev. Lett.}, 81:1359--1362, 1998.

\bibitem{Bertuzzo:2024eds}
Enrico Bertuzzo and Michele Frigerio.
\newblock {Two portals to GeV sterile neutrinos: Dipole versus mixing}.
\newblock {\em SciPost Phys.}, 18(4):140, 2025.

\bibitem{Brdar:2020quo}
Vedran Brdar, Admir Greljo, Joachim Kopp, and Toby Opferkuch.
\newblock {The Neutrino Magnetic Moment Portal: Cosmology, Astrophysics, and
  Direct Detection}.
\newblock {\em JCAP}, 01:039, 2021.

\bibitem{Magill:2018jla}
Gabriel Magill, Ryan Plestid, Maxim Pospelov, and Yu-Dai Tsai.
\newblock {Dipole Portal to Heavy Neutral Leptons}.
\newblock {\em Phys. Rev. D}, 98(11):115015, 2018.

\bibitem{Delgado:2022fea}
F.~Delgado, L.~Duarte, J.~Jones-Perez, C.~Manrique-Chavil, and S.~Pe{\~n}a.
\newblock {Assessment of the dimension-5 seesaw portal and impact of exotic
  Higgs decays on non-pointing photon searches}.
\newblock {\em JHEP}, 09:079, 2022.

\bibitem{Zhang:2023nxy}
Yu~Zhang and Wei Liu.
\newblock {Probing active-sterile neutrino transition magnetic moments at LEP
  and CEPC}.
\newblock {\em Phys. Rev. D}, 107(9):095031, 2023.

\bibitem{Jodlowski:2020vhr}
Krzysztof Jod{\l}owski and Sebastian Trojanowski.
\newblock {Neutrino beam-dump experiment with FASER at the LHC}.
\newblock {\em JHEP}, 05:191, 2021.

\bibitem{Duarte:2023tdw}
L.~Duarte, J.~Jones-P{\'e}rez, and C.~Manrique-Chavil.
\newblock {Bounding the Dimension-5 Seesaw Portal with non-pointing photon
  searches}.
\newblock {\em JHEP}, 04:133, 2024.

\bibitem{Ovchynnikov:2023wgg}
Maksym Ovchynnikov and Jing-Yu Zhu.
\newblock {Search for the dipole portal of heavy neutral leptons at future
  colliders}.
\newblock {\em JHEP}, 07:039, 2023.

\bibitem{Beltran:2024twr}
Rebeca Beltr\'an, Patrick~D. Bolton, Frank~F. Deppisch, Chandan Hati, and
  Martin Hirsch.
\newblock {Probing heavy neutrino magnetic moments at the LHC using long-lived
  particle searches}.
\newblock {\em JHEP}, 07:153, 2024.

\bibitem{Barducci:2024kig}
Daniele Barducci and Alessandro Dondarini.
\newblock {Neutrino dipole portal at a high energy
  \ensuremath{\mu}\ensuremath{-}collider}.
\newblock {\em JHEP}, 10:165, 2024.

\bibitem{Brdar:2025iua}
Vedran Brdar, Ying-Ying Li, Samiur~R. Mir, and Yi-Lin Wang.
\newblock {Collider Prospects for the Neutrino Magnetic Moment Portal}.
\newblock 2 2025.

\bibitem{FCC:2025lpp}
M.~Benedikt et~al.
\newblock {Future Circular Collider Feasibility Study Report: Volume 1,
  Physics, Experiments, Detectors}.
\newblock 4 2025.

\bibitem{FCC:2018byv}
A.~Abada et~al.
\newblock {FCC Physics Opportunities}: {Future Circular Collider Conceptual
  Design Report Volume 1}.
\newblock {\em Eur. Phys. J. C}, 79(6):474, 2019.

\bibitem{Accettura:2023ked}
Carlotta Accettura et~al.
\newblock {Towards a muon collider}.
\newblock {\em Eur. Phys. J. C}, 83(9):864, 2023.
\newblock [Erratum: Eur.Phys.J.C 84, 36 (2024)].

\bibitem{FCC:2018vvp}
A.~Abada et~al.
\newblock {FCC-hh: The Hadron Collider}: {Future Circular Collider Conceptual
  Design Report Volume 3}.
\newblock {\em Eur. Phys. J. ST}, 228(4):755--1107, 2019.

\bibitem{Frigerio:2024jlh}
Michele Frigerio and Natascia Vignaroli.
\newblock {Muon collider probes of Majorana neutrino dipole moments and
  masses}.
\newblock {\em JHEP}, 04:008, 2025.

\bibitem{Dehghani:2025xkd}
Parham Dehghani, Mariana Frank, and Benjamin Fuks.
\newblock {Vector Boson Fusion Signatures of Superheavy Majorana Neutrinos at
  Muon Colliders}.
\newblock 6 2025.

\bibitem{Balaji:2020oig}
Shyam Balaji, Maura Ramirez-Quezada, and Ye-Ling Zhou.
\newblock {CP violation in neutral lepton transition dipole moment}.
\newblock {\em JHEP}, 12:090, 2020.

\bibitem{Alwall:2014hca}
J.~Alwall, R.~Frederix, S.~Frixione, V.~Hirschi, F.~Maltoni, O.~Mattelaer,
  H.~S. Shao, T.~Stelzer, P.~Torrielli, and M.~Zaro.
\newblock {The automated computation of tree-level and next-to-leading order
  differential cross sections, and their matching to parton shower
  simulations}.
\newblock {\em JHEP}, 07:079, 2014.

\bibitem{Alloul:2013bka}
Adam Alloul, Neil~D. Christensen, C\'eline Degrande, Claude Duhr, and Benjamin
  Fuks.
\newblock {FeynRules 2.0 - A complete toolbox for tree-level phenomenology}.
\newblock {\em Comput. Phys. Commun.}, 185:2250--2300, 2014.

\bibitem{Bierlich:2022pfr}
Christian Bierlich et~al.
\newblock {A comprehensive guide to the physics and usage of PYTHIA 8.3}.
\newblock {\em SciPost Phys. Codeb.}, 2022:8, 2022.

\bibitem{Cacciari:2011ma}
Matteo Cacciari, Gavin~P. Salam, and Gregory Soyez.
\newblock {FastJet User Manual}.
\newblock {\em Eur. Phys. J. C}, 72:1896, 2012.

\bibitem{deFavereau:2013fsa}
J.~de~Favereau, C.~Delaere, P.~Demin, A.~Giammanco, V.~Lema\^\i{}tre,
  A.~Mertens, and M.~Selvaggi.
\newblock {DELPHES 3, A modular framework for fast simulation of a generic
  collider experiment}.
\newblock {\em JHEP}, 02:057, 2014.

\bibitem{ATLAS:2019jvq}
Morad Aaboud et~al.
\newblock {Electron reconstruction and identification in the ATLAS experiment
  using the 2015 and 2016 LHC proton-proton collision data at $\sqrt{s}$ = 13
  TeV}.
\newblock {\em Eur. Phys. J. C}, 79(8):639, 2019.

\bibitem{Mohan:2015doa}
Kirtimaan Mohan and Natascia Vignaroli.
\newblock {Vector resonances in weak-boson-fusion at future pp colliders}.
\newblock {\em JHEP}, 10:031, 2015.

\bibitem{Goncalves:2017gzy}
Dorival Goncalves, Tilman Plehn, and Jennifer~M. Thompson.
\newblock {Weak boson fusion at 100 TeV}.
\newblock {\em Phys. Rev. D}, 95(9):095011, 2017.

\bibitem{Baker:2022zxv}
Michael~J. Baker, Timothy Martonhelyi, Andrea Thamm, and Riccardo Torre.
\newblock {The role of vector boson fusion in the production of heavy vector
  triplets at the LHC and HL-LHC}.
\newblock {\em JHEP}, 11:066, 2022.

\bibitem{Molinaro:2017mwb}
Emiliano Molinaro, Francesco Sannino, Anders~Eller Thomsen, and Natascia
  Vignaroli.
\newblock {Uncovering new strong dynamics via topological interactions at the
  100 TeV collider}.
\newblock {\em Phys. Rev. D}, 96(7):075040, 2017.

\bibitem{Costantini:2020stv}
Antonio Costantini, Federico De~Lillo, Fabio Maltoni, Luca Mantani, Olivier
  Mattelaer, Richard Ruiz, and Xiaoran Zhao.
\newblock {Vector boson fusion at multi-TeV muon colliders}.
\newblock {\em JHEP}, 09:080, 2020.

\bibitem{Vignaroli:2023rxr}
Natascia Vignaroli.
\newblock {Charged resonances and MDM bound states at a multi-TeV muon
  collider}.
\newblock {\em JHEP}, 10:121, 2023.

\end{thebibliography}

\end{document}